\pdfoutput=1
\documentclass[fleqn,usenatbib,useAMS]{mnras}

\usepackage{amssymb}	
\usepackage{multicol}        
\usepackage{bm}		
\usepackage{pdflscape}	
\usepackage[T1]{fontenc}
\usepackage{ae,aecompl}

\usepackage {graphicx}
\usepackage{natbib}
\usepackage{aas_journals}
\usepackage{color}
\usepackage{amsmath}
\definecolor{mygray}{gray}{0.7}

\newcommand{\unit}[1]{\textnormal{#1}}
\newcommand{\Hunit}{\ensuremath{\frac{\unit{km}/\unit{s}}{\unit{Mpc}}}}

\def\aap{{ A\&A}}

\def\apj{{ApJ}}
\def\apjl{{ApJL}}

\def\mnras{{MNRAS}}

\def\jcap{JCAP}

\def\nat{Nature}

\def\apjs{{ApJS}}

\newcommand{\bes}{\begin{equation*}}
\newcommand{\ees}{\end{equation*}}
\newcommand{\bea}{\begin{eqnarray}}
\newcommand{\eea}{\end{eqnarray}}
\newcommand{\beas}{\begin{eqnarray*}}
\newcommand{\eeas}{\end{eqnarray*}}

\newcommand{\mpc}{\rm {h^{-1}Mpc }}

\newcommand{\ltsima}{$\; \buildrel < \over \sim \;$}
\newcommand{\lsim}{\lower.5ex\hbox{\ltsima}}
\newcommand{\gtsima}{$\; \buildrel > \over \sim \;$}
\newcommand{\gsim}{\lower.5ex\hbox{\gtsima}}

\def\gtrsim{\mathrel{\hbox{\rlap{\hbox{\lower4pt\hbox{$\sim$}}}\hbox{$>$}}}}
\let\ga=\gtrsim
\def\lesssim{\mathrel{\hbox{\rlap{\hbox{\lower4pt\hbox{$\sim$}}}\hbox{$<$}}}}
\let\la=\lesssim

\definecolor{mygray}{gray}{0.5}

\newcommand{\be}{\begin{equation}}
\newcommand{\ee}{\end{equation}}
\newcommand{\ba}{\begin{eqnarray}}
\newcommand{\ea}{\end{eqnarray}}

\title[ISW imprint of supervoids as a probe of dark energy]{A common explanation of the Hubble tension and anomalous cold spots in the CMB}

\author[Andr\'as Kov\'acs et al.]{A. Kov\'acs$^{1,2}$\thanks{Juan de la Cierva Fellow, Email: akovacs@iac.es}, R. Beck$^{3,4}$, I. Szapudi$^3$, I. Csabai$^4$, G. R\'acz$^4$, L. Dobos$^4$\\
$^{1}$ Instituto de Astrof\'{\i}sica de Canarias (IAC), Calle V\'{\i}a L\'{a}ctea, E-38200, La Laguna, Tenerife, Spain\\
$^{2}$ Departamento de Astrof\'{\i}sica, Universidad de La Laguna (ULL), E-38206, La Laguna, Tenerife, Spain\\
$^{3}$ Institute for Astronomy, University of Hawaii, 2680 Woodlawn Drive, Honolulu, HI, 96822 \\
$^{4}$ Department of Physics of Complex Systems, ELTE E\"{o}tv\"{o}s Lor\'and University, Pf. 32, H-1518 Budapest, Hungary}

\begin{document}
\date{Submitted 2020}
\pagerange{\pageref{firstpage}--\pageref{lastpage}} \pubyear{2017}
\maketitle
\label{firstpage}
\begin{abstract}
The standard cosmological paradigm narrates a reassuring story of a universe currently dominated by an enigmatic dark energy component. Disquietingly, its universal explaining power has recently been challenged by, above all, the $\sim4\sigma$ tension in the values of the Hubble constant. Another, less studied anomaly is the repeated observation of integrated Sachs-Wolfe imprints $\sim5\times$ stronger than expected in the $\Lambda$CDM model from $R_{\rm v}\gsim100\mpc$ super-structures. Here we show that the inhomogeneous AvERA model of emerging curvature is capable of telling a plausible albeit radically different story that explains both observational anomalies \emph{without dark energy}. We demonstrate that while stacked imprints of $R_{\rm v}\gsim100\mpc$ supervoids in cosmic microwave background temperature maps can discriminate between the AvERA and $\Lambda$CDM models, their characteristic differences may remain hidden using alternative void definitions and stacking methodologies. Testing the extremes, we then also show that the CMB Cold Spot can plausibly be explained in the AvERA model as an ISW imprint. The coldest spot in the AvERA map is aligned with multiple low-$z$ supervoids with $R_{\rm v}\gsim100\mpc$ and central underdensity $\delta_{0}\approx-0.3$, resembling the observed large-scale galaxy density field in the Cold Spot area. We hence conclude that the anomalous imprint of supervoids may well be the canary in the coal mine, and existing observational evidence for dark energy should be re-interpreted to further test alternative models.

\end{abstract}

\section{Introduction}

A key puzzle in modern cosmology is the nature and origin of cosmic acceleration in the late-time Universe. The concordance $\Lambda$-Cold Dark Matter ($\Lambda$CDM) model has shown remarkable success in fitting various observational results and it tells a successful yet counter-intuitive story of a universe filled mostly with dark energy. At present, constraints on the model parameters are becoming so precise that cosmologists have seemingly painted themselves into a corner with very limited room for manoeuvring. 

Above all, an intriguing tension has emerged in the determination of the Hubble constant. Based on the cosmic distance ladder method using Cepheids and supernovae, the latest estimate by \cite{Riess2019} is $H_0=74.03\pm1.42~\Hunit$. In contrast, the extrapolated value from analyses of the cosmic microwave background (CMB) \citep{Planck2018} \emph{assuming} a $\Lambda$CDM model is $H_0=67.5\pm0.5~\Hunit$, i.e. the discrepancy is over $4.4\sigma$ \citep{Verde2019}. While proposed solutions may come from modifications in the concordance model at late or early times, the majority of the candidate explanations assume new degrees of freedom in the early Universe to change the sound horizon \citep[see e.g.][]{Bernal2016,Knox2019}.

\begin{figure*}
\begin{center}
\includegraphics[width=170mm]{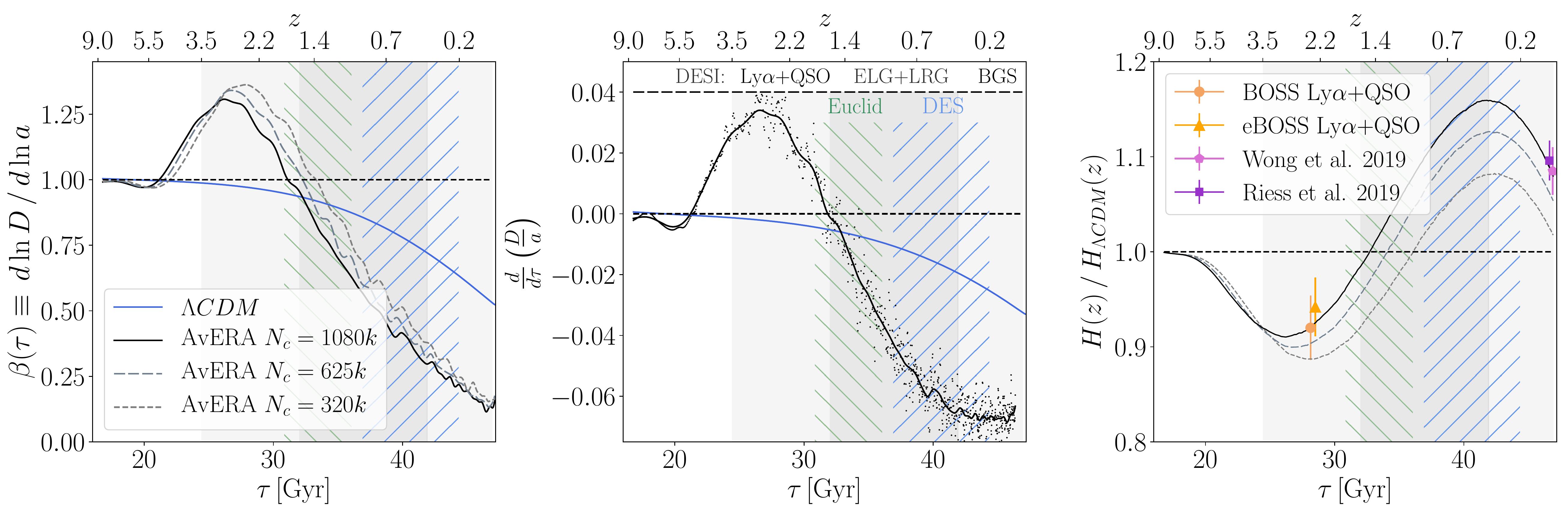}
\caption{\label{fig:figure_1}Cosmological growth factor ($D$) derivatives (left, middle) and expansion history functions (right) utilised in ISW calculations ($a$ is the scale factor while $\tau$ represents conformal time). We compare three AvERA simulation runs with different coarse graining scales (black and gray dashed curves) and the corresponding values for a standard $\Lambda \mathrm{CDM}$ model (blue) with the best-fit parameters from \emph{Planck} (points in the middle panel are numerical derivatives, lines are with Gaussian smoothing applied). Vertical bands mark observational redshift windows by current (DES) and future (DESI, Euclid) galaxy surveys that may be used to probe AvERA and similar models. In expansion history relative to that of $\Lambda \mathrm{CDM}$, the right panel highlights the explaining power of AvERA in observed $H_0$ and $H(z\approx2.34)$ anomalies at low and intermediate redshifts.}
\end{center}
\end{figure*}

\subsection{The AvERA model and the $H_0$ tension}

The AvERA (Average Expansion Rate Approximation) inhomogeneous cosmological simulation by \cite{Racz2017} offered a novel late late-time solution that naturally reconciled the tensions between CMB-based extrapolated $H_0$ values and more direct estimates from the nearby Universe. A twist in AvERA's Newtonian N-body simulation technique is the reversed order of volume-averaging and expansion rate computation. In a \emph{modified} $\Omega_{\rm M}=1$ Einstein--de Sitter (EdS) setting, the expansion rate of smaller regions is determined by their own local density via the Friedmann equation. In turn, the effective scale factor of the simulated AvERA universe is determined based on a volume average of the locally measured scale factor increments. This procedure results in a late time effective $\Omega_{M}^{\rm eff}\approx0.26$ matter density parameter that is consistent with its $\Lambda$CDM equivalent.

Formally, the AvERA model represents a heuristic implementation of the emerging curvature class of models \citep[see e.g.][]{Bolejko2018,Heinesen2020} in which negative curvature emerges as cosmic voids begin to dominate the volume of the universe through non-linear evolution at late times. With a single adjustable parameter to set the coarse graining scale of cells for averaging, the AvERA model automatically results in a very similar expansion history to that of $\Lambda$CDM. The evolution of the AvERA model is equivalent to the case of a global Friedmann equation in the limit of large coarse graining scales where the matter distribution is approximately homogeneous. The related AvERA simulation products are publicly available\footnote{https://github.com/eltevo/avera}.

The AvERA model does not challenge the appearance of cosmic acceleration and it does not claim that we observe the Universe from the centre of a large under-dense region \citep[see e.g.][]{Kenworthy2019}. Instead, it provides an alternative mechanism that does not require a dark energy component, but is nonetheless consistent with the $\Lambda$CDM expansion history given observational errors.
Uncommonly however, the AvERA model does allow \emph{both} early and late measurements of $H_{\rm 0}$ to be correct due to characteristic small deviations from $\Lambda$CDM expansion history. 

In Figure \ref{fig:figure_1}, we provide a conceptual explanation for the alleviation of the Hubble tension in AvERA. In matter domination at $1.5<z<4.4$, the absence of the cosmological constant ($\Lambda$) term in AvERA's initial EdS-like setting results in faster structure formation, and a lower effective expansion rate $H(z)$ emerges from the AvERA model compared to $\Lambda$CDM. We note that this difference is consistent with the $\sim2\sigma$ tension in the low $H(z)$ value extracted from analyses of the baryonic acoustic oscillation (BAO) feature in quasar data at $z\approx2.34$ \citep[see e.g.][]{BOSS_LyA_BAO,Agathe2019} but it remains to be tested if AvERA can resolve this tension in detail.

The trend then changes at lower redshifts ($z<1.5$) and AvERA predicts \emph{consistently larger} $H(z)$, including a higher $H_0$ value, than $\Lambda$CDM due to the ever increasing contribution of faster expanding void regions to the volume of the universe. In Figure \ref{fig:figure_1}, we compare observational $H_0$ constraints including H0LiCOW ($H_0$ Lenses in COSMOGRAIL's Wellspring) results \citep{Wong2019} and three different AvERA model versions, as a function of the number of cells used for averaging. As reported by \cite{Racz2017} using their $V_{\rm sim}\approx(100\,\mpc)^{3}$ simulation, coarse graining using $N_{c}$=1,080,000 cells and a corresponding $1.17 \times 10^{11} \, M_\odot$ mean particle mass in $V_{c}\approx1\, (\mpc)^{3}$ cubic cells results in the best agreement between observations of $H_{\rm 0}$ and $H(z)$ and their values in AvERA.

\subsection{ISW anomalies interpreted in AvERA}

The AvERA model also predicts characteristic differences in the growth rate of structure as a function of redshift compared to $\Lambda$CDM. A prime observable related to growth history, argued \cite{Beck2018}, to distinguish between the two cosmologies is the integrated Sachs-Wolfe effect \citep[ISW]{Sachs1967,Rees1968}. The ISW signal is sourced by a stretching effect due to the late-time imbalance of structure growth and cosmic expansion, i.e. cosmic acceleration. As large-scale gravitational potentials decay along the paths of CMB photons at late times, tiny secondary CMB temperature anisotropies are imprinted on the primary CMB temperature fluctuations. Therefore, details of cosmic acceleration and the growth rate of structure can be studied using CMB photons that directly probe the largest cosmic structures (hosted by the changing potentials) as they experience space-stretching effects.

Notably, observations of the ISW effect too have their own anomalies. The CMB imprint of the largest structures in the cosmic web at $R_{\rm v}\gsim100\mpc$ scales appears to be about $\sim4-5\times$ stronger than expected in the $\Lambda$CDM model \citep[see e.g.][]{Granett2008,Kovacs2019}. Inconsistently though, measurements using full two-point cross-correlation or other void catalogues containing smaller voids showed no significant excess even from the very same galaxy survey data \citep[see e.g.][]{PlanckISW2015, NadathurCrittenden2016, Stolzner2018}. 

A related problem is the alignment of \emph{the} Cold Spot anomaly in the CMB \citep[see e.g.][]{CruzEtal2004} and the low-$z$ Eridanus supervoid \citep[see e.g.][]{SzapudiEtAl2014}. While the ISW imprint of such a supervoid is not strong enough to explain the temperature profile of the Cold Spot in a $\Lambda$CDM model \citep[see e.g.][]{Nadathur2014}, this case may be related to the ISW puzzle if further evidence is presented for excess ISW signals.

Providing a basis to test these claims in the AvERA framework, \cite{Beck2018} created a reconstructed ISW sky map from an N-body simulation via ray-tracing techniques. Their auto-correlation analysis of the resulting simulated ISW map showed that the AvERA model predicts about $\sim2-5\times$ times larger ISW signals depending on the angular scales. In Figure \ref{fig:figure_1}, we demonstrate that the AvERA model predicts not only higher expansion rate at low redshifts, as discussed above, but also larger values for the derivatives of the $D$ linear growth factor compared to $\Lambda$CDM. Once voids begin to dominate the effective expansion rate of the universe at $z<1.5$, gravitational potentials decay \emph{faster} than in $\Lambda$CDM scenarios, resulting in stronger ISW temperature imprints for large-scale structures.

We note that the choice of $N_c$, that sets cell size for volume averaging, does affect the AvERA predictions for the growth rate of structure, but all AvERA model versions predict larger amplitude for the ISW effect than its $\Lambda$CDM equivalent. Since a coarse graining with $V_{c}\approx1\, (\mpc)^{3}$ cell size appears to outperform other AvERA versions in explaining observational anomalies of expansion history, we are guided to focus on tests of the ISW predictions of this particular setup.

\subsection{Detecting the ISW imprint of cosmic voids}

The ISW effect in either model, however, is too weak to be discernible  in auto-correlation analyses of CMB temperature maps. The CMB data should rather be \emph{cross}-correlated with tracer catalogues of the matter density fluctuations in the late Universe \citep{CrittendenTurok1996}. In this paper, we implement such a cross-correlation to complement previous analyses by \cite{Beck2018}. 

The ray-traced ISW map necessarily contains additional information beyond what is accessible by two-point analyses. Higher-order statistics or simple visual inspections of features may help accessing this extra information \citep[see e.g.][and references therein]{Kitaura2016}. We therefore identify the largest under-dense regions in the cosmic web (the so-called cosmic voids and supervoids) with the strongest expected ISW signal, and then, instead of formal two-point analyses, we perform a \emph{stacking} analysis of CMB patches aligned with them to understand the anomalous measurements that followed the same strategy \citep[see e.g.][and references therein]{Kovacs2019}. We also identify the \emph{coldest} spots in the resulting AvERA maps to test the most extreme cases.

To further motivate our analyses, we highlight in Figure \ref{fig:figure_1} that the $0.2 < z < 0.9$ redshift window of the Dark Energy Survey (DES), i.e. where ISW excess signals have been seen, appears to be almost ideal for studying the characteristic differences between the AvERA and $\Lambda$CDM cosmologies. 

In a broader context, an interesting feature of the AvERA cosmology is the slower EdS-like expansion rate, and the corresponding faster gravitational growth and positive growth factor derivative at $1.5<z<4.4$, before cosmic voids begin their dominance, as shown in Figure \ref{fig:figure_1}. Such effects, sourced by the growth of the gravitational potentials at high redshifts in the absence of a $\Lambda$ component, are \emph{not} predicted by the $\Lambda$CDM model where the consequence of the dominating $\Lambda$ term at late times is an increasingly stronger decay of the potentials. This faster high-$z$ growth rate, compensated at low redshifts by a suppressed growth that is even stronger than in $\Lambda$CDM, implies an opposite-sign ISW effect from the $1.5<z<4.4$ redshift range. Future data from the Dark Energy Spectroscopic Instrument (DESI) \citep{DESI} may probe in great detail using quasar tracers (QSO) or the Lyman-$\alpha$ (Ly$\alpha$) forest at the peak of the quasar space density. At $0.4 < z < 1.6$, emission line galaxies (ELG) and luminous red galaxies (LRG) in DESI data may be used to follow up on DES results, and potentially extend the analysis to the lowest redshifts with the Bright Galaxy Survey (BGS). We also find, however, that the Euclid redshift survey \citep{euclid} will most probably be less sensitive to the AvERA vs. $\Lambda$CDM differences at $1.0<z<1.8$.

As introduced below in Section 2, we base our analysis on identifying cosmic voids of various types in a tracer catalogue extracted from a large N-body simulation. In Section 3, we describe our methodologies to define and identify cosmic voids and how to estimate their mean imprint on the CMB using a stacking cross-correlation technique. The most important findings of the paper are presented in Section 4 where we provide a detailed analysis of the ISW imprints of (super)voids in $\Lambda$CDM and AvERA models, followed by Section 5 that presents our results on the extreme case of the CMB Cold Spot. We then discuss our findings in Section 6, and conclude in Section 7 that the AvERA model may not only solve the $H_0$ tension but the ISW puzzle as well. 

\section{Data sets for cross-correlations}

The stacking cross-correlation measurement we wish to perform requires a catalogue of cosmic voids \emph{and} a reconstructed ISW map from the same simulation. We based our analysis on the Millennium-XXL (MXXL) dark matter only $\Lambda$CDM N-body simulation by \cite{Angulo2012}. The MXXL is an upgraded version of the earlier Millennium run \citep{Springel2005}, covering a co-moving volume of ($3h^{-1}$ Gpc)$^{3}$ with $6720^3$ particles of mass $8.456 \times 10^9 \, M_\odot$. It adopts cosmological parameters consistent with the WMAP-1 mission results \citep{Spergel2003}.

We note that the value of the matter density parameter $\Omega_{\rm m}=0.25$ applied in MXXL is lower than the more modern result from the {\it Planck} survey \citep{Planck2018} with $\Omega_{\rm m}=0.315\pm0.007$. Relevantly, however, \cite{Nadathur2012} reported that differences in modelled ISW imprints of super-structures are significantly smaller for values $0.25<\Omega_{\rm m}<0.32$ than the error bars and the level of excess signals from observations.

An important aspect is that the $\Lambda$CDM expansion history of the MXXL simulation is slightly different than that of AvERA. Therefore, the MXXL matter density field, and its large-scale features such as supervoids, cannot be directly paired to AvERA cosmological parameters throughout cosmic history. \cite{Beck2018} argued that, as inhomogeneities $z\approx9$ are still small and different cosmologies match closely, using such an early snapshot of the MXXL simulation for the AvERA model is a reasonable solution. The linear growth approximation is then applied to evolve the density field from this less evolved early state following the AvERA growth function. We assume that the supervoids we wish to study are well described by linear theory because they trace extended ($\gsim100$ $\mpc$) but shallow ($\bar{\delta}_{\rm m}\sim-0.1$ mean internal density contrast, with $\delta=\rho/\bar{\rho}-1$) fluctuations in the matter density field (see further details in Figure \ref{fig:figure_3} below). We also note that supervoids are originated from the most extended negative primordial fluctuations in the matter density field. Consequently, their size and magnitude is expected to be similar in both models.

\subsection{A mock catalogue of tracers}

We use the publicly available full sky MXXL halo light-cone catalogue by \cite{Smith2017} to facilitate our ISW-density cross-correlation measurement. We do not apply any mask in our analysis since full sky data provides more precise estimates of the true ISW signal of supervoids.

Another important parameter is the redshift cut. The light-cone mock extends to $z= 2.2$ yet we only select tracers of redshifts $z<0.9$ where the stacked ISW signal of cosmic voids has been measured using Baryon Oscillation Spectroscopic Survey (BOSS) and DES galaxy survey data \citep[see][for more details]{Kovacs2019}. 

As a further refinement, we apply a halo mass cut in order to approximately model the population of luminous red galaxies (LRG) that were used as tracers of supervoids in observed data. Typically, LRGs are expected to reside in halos of mass $\sim10^{13}-10^{14}h^{-1} M_\odot$ \citep[see e.g.][]{Zheng2009,Hotchkiss2015} well above the mass resolution of the MXXL halo mock catalogue with $\sim10^{11}h^{-1} M_\odot$. We thus apply a simple halo mass cut with $M_{\rm 200m} >10^{13}h^{-1}M_\odot$ to define an LRG-like population. The corresponding tracer density, according to \cite{Smith2017}, changes from $\bar{n}\approx3\times10^{-4}h^{3}$ Mpc$^{-3}$ to $\bar{n}\approx 10^{-4}h^{3}$ Mpc$^{-3}$ from redshift $z=0$ to $z=1$. This displays close agreement with the tracer density of the observed DES Year-3 LRG sample that we wish to model in this analysis \citep[see][]{Kovacs2019}. 

We therefore query the MXXL light-cone data base\footnote{https://tao.asvo.org.au/tao/} and extract the following data columns: Right Ascension (RA), Declination (Dec), cosmological redshift $z_{\rm cos}$, and halo mass $M_{\rm 200m}$ in units of $10^{10}h^{-1}M_\odot$. 

\subsection{Simulated ISW temperature maps}

A standard reconstruction method of ISW maps is ray-tracing \cite[see e.g.][]{Cai2010}. Recently, \cite{Beck2018} created ray-traced ISW maps by following the path of light-rays through the MXXL simulation (from random starting points). This simulation is large enough to be suitable for ISW analyses since the light-ray reaches the boundary of the simulation box for the first time at a redshift of $z \approx 1.3$. Therefore, no significant effects are expected from the application of periodic boundary conditions as most of the ISW signal is expected to come from $z\lsim1$ in both cosmological models we consider.

\begin{figure*}
\begin{center}
\includegraphics[width=85mm]{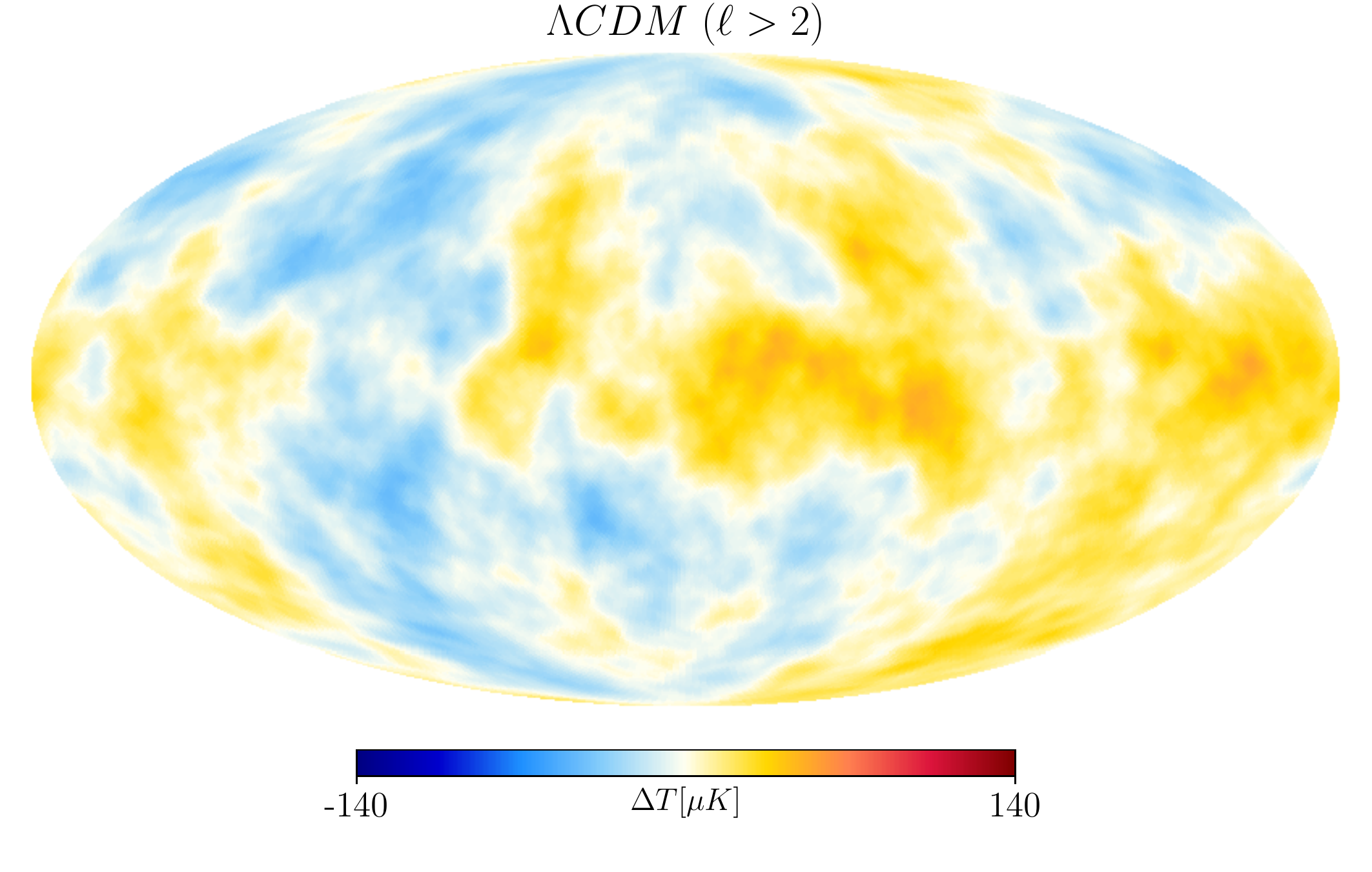}\includegraphics[width=85mm]{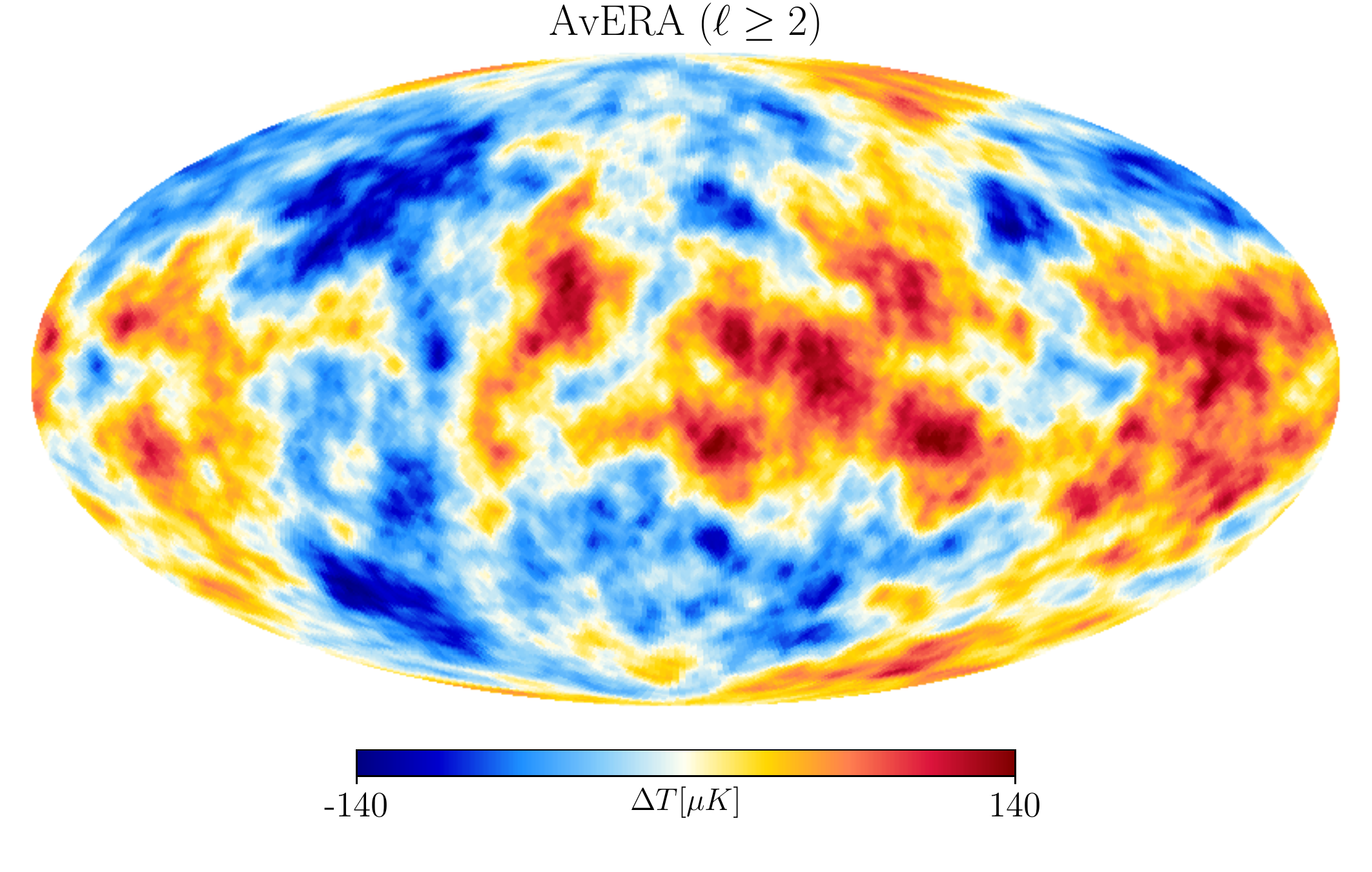}\\
\includegraphics[width=85mm]{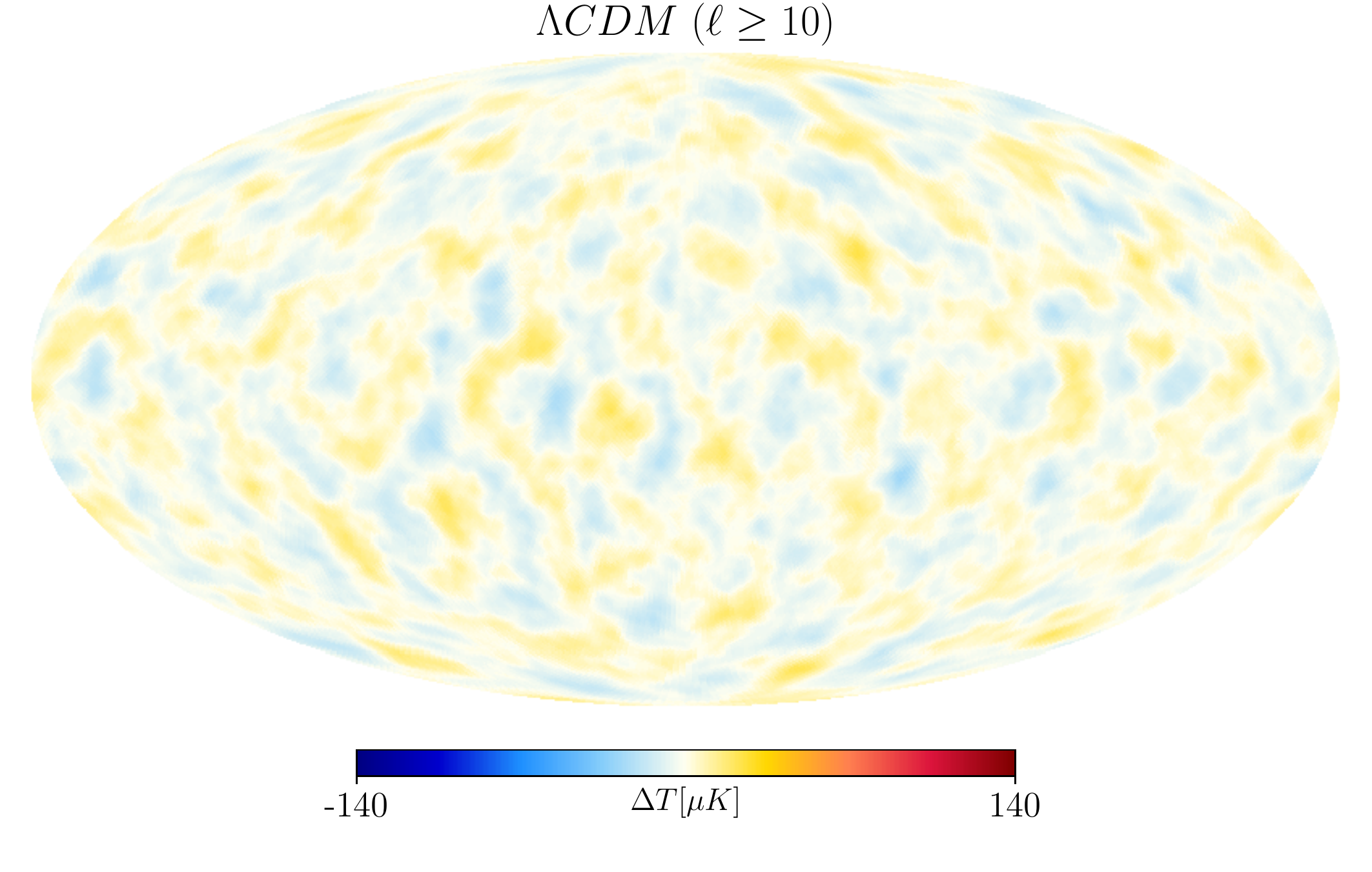}\includegraphics[width=85mm]{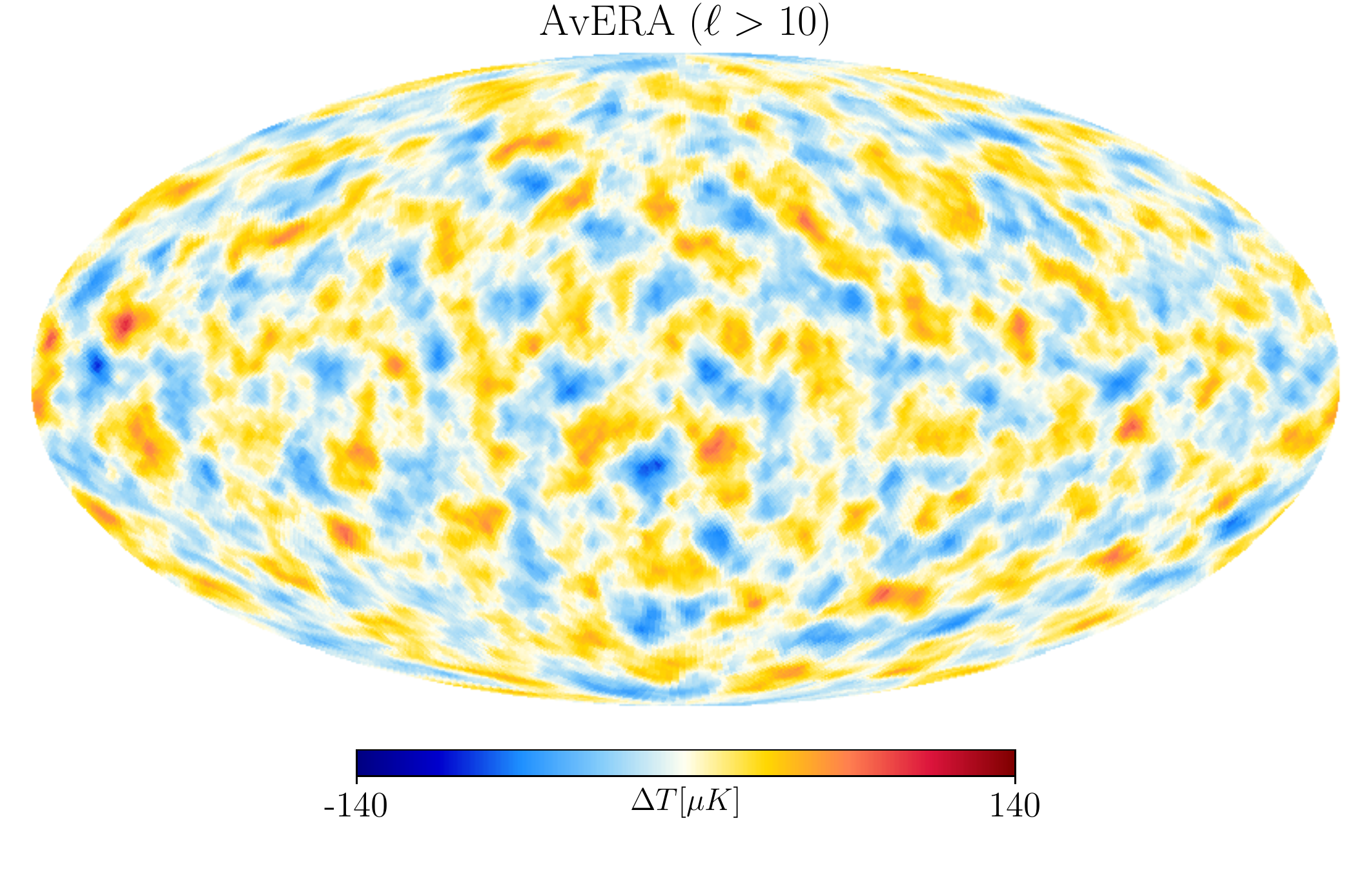}
\caption{\label{fig:figure_2}Ray-traced ISW temperature maps are compared using $\Lambda$CDM (left) and AvERA (right) cosmological models. The top panels show the maps with all available modes included while the bottom panels illustrate the remaining temperature fluctuations without the largest modes $2\leq \ell \leq10$ that add extra noise to the stacked signal, or may bias the signal if measured in smaller patches. The use of a common colour scale highlights the differences in the magnitudes of the residual fluctuations.}
\end{center}
\end{figure*}

We use the publicly available\footnote{https://github.com/beckrob/AvERA\_ISW} ISW reconstruction code by \cite{Beck2018} to produce the two ISW maps from MXXL assuming $\Lambda$CDM and AvERA models. As starting point for the ray-tracing, we choose that of the halo lightcone catalogue. The resulting \texttt{HEALPix} \citep{healpix} maps show very similar large-scale features, as shown in Figure \ref{fig:figure_2}. The fluctuations in the AvERA ISW map, however, are of higher amplitude as expected based on previous results. The reason for such a difference is the larger low-redshift derivative of the growth function compared to $\Lambda$CDM that we presented in Figure \ref{fig:figure_1}. We note that the AvERA map is not simply a re-scaled version of the $\Lambda$CDM map because structures at given redshifts contribute differently in the two models. At $z>1.4$, for instance, AvERA predicts an ISW signal of opposite sign and higher amplitude compared to $\Lambda$CDM. This signal may result in specific cancellation and mixing effects that may also alter the detectability of the low-$z$ AvERA signal as extra noise. For further details about differences between models see Figure \ref{fig:figure_1}, while a detailed description of the ray-tracing methodology is given in Appendix \ref{app_raytracing}.

Additionally, we test the effects of low-$\ell$ modes in the ISW maps. We follow \cite{Kovacs2019} who reported that large-scale modes add extra noise to the stacked profile and potentially introduce biases in the measured ISW profiles if measured in smaller patches. We therefore remove the contributions from the largest modes with multipoles $2\leq \ell \leq10$. This procedure results in a lack of irregular large-scale patches and lower uncertainties in the stacking measurement, with a slight reduction in the stacked signal's amplitude.

\section{Methods}

We use the resulting tracer catalogue of $12,486,811$ high-mass MXXL halos to identify cosmic voids. We highlight that observational evidence of the ISW anomaly suggests that details of the void finding may be crucially important to the problem. Cosmic voids are highly hierarchical objects in the cosmic web with two main classes. Voids-in-clouds tend to be surrounded by an over-dense environment, while voids-in-voids, or supervoids, consist of several sub-voids \citep[see e.g.][]{Sheth2004,Lares2017}.

Such large-scale void structures are of high interest in ISW measurements, as they are expected to account for most of the observable cold spot signal. In particular, \cite{Kovacs2018} found that large $R_{\rm v}\gsim100$ $\mpc$ supervoids imprint a specific ISW pattern with a central cold spot and a surrounding hot ring. Importantly, anomalous ISW-like signals are typically seen if merging of voids into larger encompassing under-densities is allowed in the void finding process \citep[see e.g.][]{Cai2017,Kovacs2018}. In contrast, no significant excess signals have been reported from the same data set when using definitions without void merging \citep[see e.g.][and references therein]{NadathurCrittenden2016}. 

Alternatively, void merging can also be a consequence of the tracer data properties themselves. In the case of photometric redshift data such as DES LRGs, extended void structures elongated in the line-of-sight are identified more efficiently because of the smearing effect of photo-$z$ errors that erase voids surrounded by over-densities \citep[see e.g.][]{Granett2015,Kovacs2016}. The photon travel time is longer through these elongated under-densities than in spherical voids of the same angular size, and thus the corresponding larger ISW temperature shifts are easier to detect. Therefore, somewhat counter-intuitively, photo-$z$ data and elongated voids can outperform voids from spec-$z$ methods in terms of $S/N$ expected from ISW or lensing measurements \citep[see e.g.][]{Cautun2018,Fang2019}. Naturally, such effects should be taken into account in simulated estimates of the ISW imprints of such voids following \cite{Kovacs2019}. While a stacked signal of larger amplitude may in general be possible to find in such an observational setup, \cite{Flender2013} concluded that large samples of elongated and spherical voids are expected to imprint statistically consistent ISW signals in a $\Lambda$CDM model.

Related to these findings, we perform new tests using our simulated AvERA and $\Lambda$CDM ISW maps and catalogues of voids and supervoids based on a stacking methodology.

\subsection{Void finding: supervoids}

Supervoids are extended but, on average, relatively shallow systems of several merged sub-voids, tracing negative fluctuations in the matter density field at the largest scales. Our main goal is to model the measurement by \cite{Kovacs2019} who used 87 supervoids identified in DES Year-3 photo-$z$ data \citep[see also][]{Sanchez2016}. This 2D void finding method is a restriction to tomographic slices of galaxy data, and analyses of the projected density field around void centre candidates defined by minima in the smoothed density field. A suitable line-of-sight slicing was found to be of thickness $\approx100~h^{-1}{\rm Mpc}$ for photo-$z$ errors at the level of $\sigma_{z}/(1+z) \approx 0.02$ or $\sim50~h^{-1}{\rm Mpc}$ at $z\approx0.5$. We model the effects of photo-$z$ smearing in our MXXL void finding procedure by adding Gaussian errors with $\sigma_z/(1+z)\approx 0.02$ to MXXL redshifts. We then slice the halo lightcone catalogue into shells of $100~h^{-1}{\rm Mpc}$.

The Gaussian smoothing applied to the tracer map is a free parameter in the process and the merging of average voids into larger supervoids may be achieved using higher smoothing scales. In practical terms, while for example a $\sigma=20~h^{-1}{\rm Mpc}$ smoothing allows one to detect more voids, as a result of void merging a $\sigma=50~h^{-1}{\rm Mpc}$ smoothing, applied also by \cite{Kovacs2019}, returns a higher number of extended $R_{\rm v}\gsim100$ $\mpc$ supervoids that we aim to study in this work.

\begin{figure}
\begin{center}
\includegraphics[width=95mm]{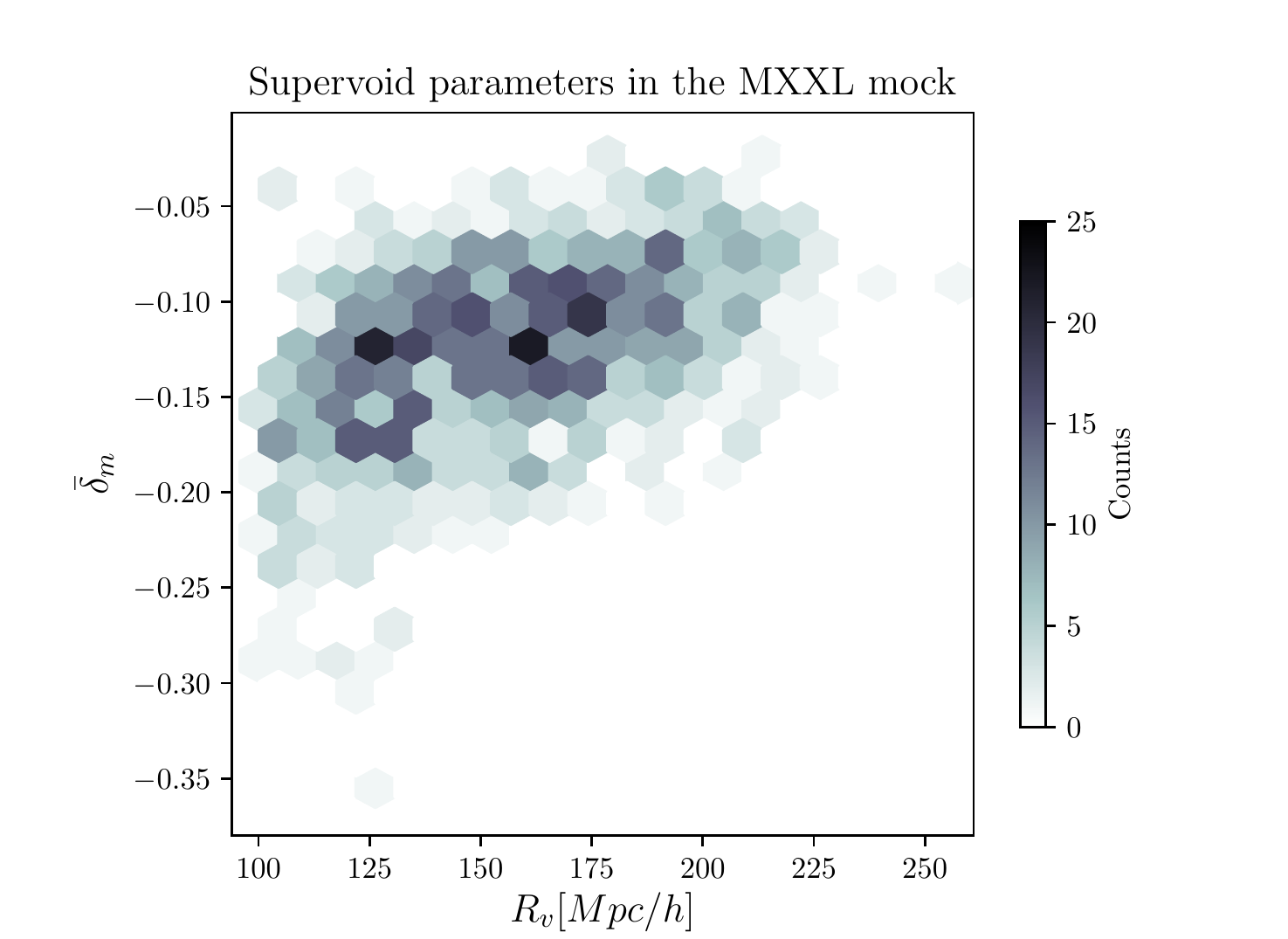}
\caption{\label{fig:figure_3}A 2-dimensional view of void parameters $\bar{\delta}_{\rm m}$ (average matter density contrast inside voids) and the $R_{\rm v}$ (void radius) using hexagonal cells.}
\end{center}
\end{figure}

With the above methodology, we identified $965$ supervoids of radii $R_{\rm v}\gsim100$ $\mpc$ at redshifts $0.2<z<0.9$ in the MXXL mock. This catalogue provides a basis for accurate estimation of the stacked ISW signals in AvERA and $\Lambda$CDM models. In Figure \ref{fig:figure_3}, we illustrate the relation of the mean under-density in voids ($\bar{\delta}_{\rm m}$) and the void radius ($R_{\rm v}$), indicating a population of shallow but extended under-densities that resemble the reported properties of observed supervoids in DES Year-3 data \citep[][]{Kovacs2019}.

\subsection{Void finding: \texttt{REVOLVER} voids}

We intended to test how the stacked imprints in AvERA and $\Lambda$CDM models differ using an alternative definition of voids. Therefore, we also run a publicly available\footnote{https://github.com/seshnadathur/REVOLVER/} void finder code on the MXXL light-cone mock catalogue. The \texttt{REVOLVER} code (REalspace VOid Locations from surVEy Reconstruction) is based on the widely used \texttt{ZOBOV} algorithm \citep{ZOBOV} and it comes in two different versions. In the version we use, first a Voronoi tessellation of the tracers is reconstructed to estimate the local density, followed by a void-finder process based on the watershed method. The centre of each void is defined as the centre of the largest empty sphere that can be placed within it. This implementation of the watershed algorithm, used for example by \cite{NadathurCrittenden2016}, prevents the merging of neighbouring voids, i.e. returns more voids that are on average relatively small.

We apply a $0.4<z<0.7$ redshift cut in the MXXL mock in order to model the measurements by \cite{NadathurCrittenden2016} who used BOSS CMASS (constant-mass) galaxies. We then further prune the resulting void catalogue as not all void types are expected to contribute with the right sign of imprint. It was found that only under-compensated voids with $\lambda_v<0$ imprint negative ISW signals, defined as
\begin{equation}
\label{eq:lambda_v}
\lambda_v\equiv\overline\delta_g\left(\frac{R_{\rmn{eff}}}{1\;h^{-1}\rmn{Mpc}}\right)^{1.2}
\end{equation}
using the average galaxy density contrast, $\overline\delta_g = \frac{1}{V}\int_{V}\delta_g\,\rmn{d}^3\mathbf{x}$, and the effective spherical radius, $R_\rmn{eff}= \left(\frac{3}{4\pi}V\right)^{1/3}$, where the volume $V$ is determined from the sum of the volumes of Voronoi cells making up the void.

We identified 46,950 voids in total at redshifts $0.4<z<0.7$ using the \texttt{REVOLVER} code. We then pruned the sample to only contain voids with $\lambda_v <0$, resulting in a sample size of 15,016 voids expected to show a cold imprint which could then be measured with stacking. The mean void size in the pruned sample is  $R_\rmn{eff}\approx43~h^{-1}\rmn{Mpc}$, indeed probing a different regime compared to supervoids.

\subsection{Stacking measurement}

Given the void parameters in the resulting catalogs, we cut out square-shaped patches from the MXXL ISW temperature maps aligned with void positions using the \texttt{gnomview} projection method of \texttt{HEALPix} \citep{healpix}. We then stacked the patches to provide a simple and informative way to statistically study the mean imprint of cosmic super-structures \cite[see e.g.][and references therein]{Kovacs2019}. 

In our main analysis, we re-scaled the cut-out patches knowing the angular size of the voids. Therefore, boundaries coincide in the stacked image as well as the void centres. In the case of \texttt{REVOLVER} voids, we also stacked the images using a fixed $\theta=15^{\circ}$ angular size in order to test the role of re-scaling in measurements by \cite{NadathurCrittenden2016}. 

\begin{figure*}
\begin{center}
\includegraphics[width=180mm]{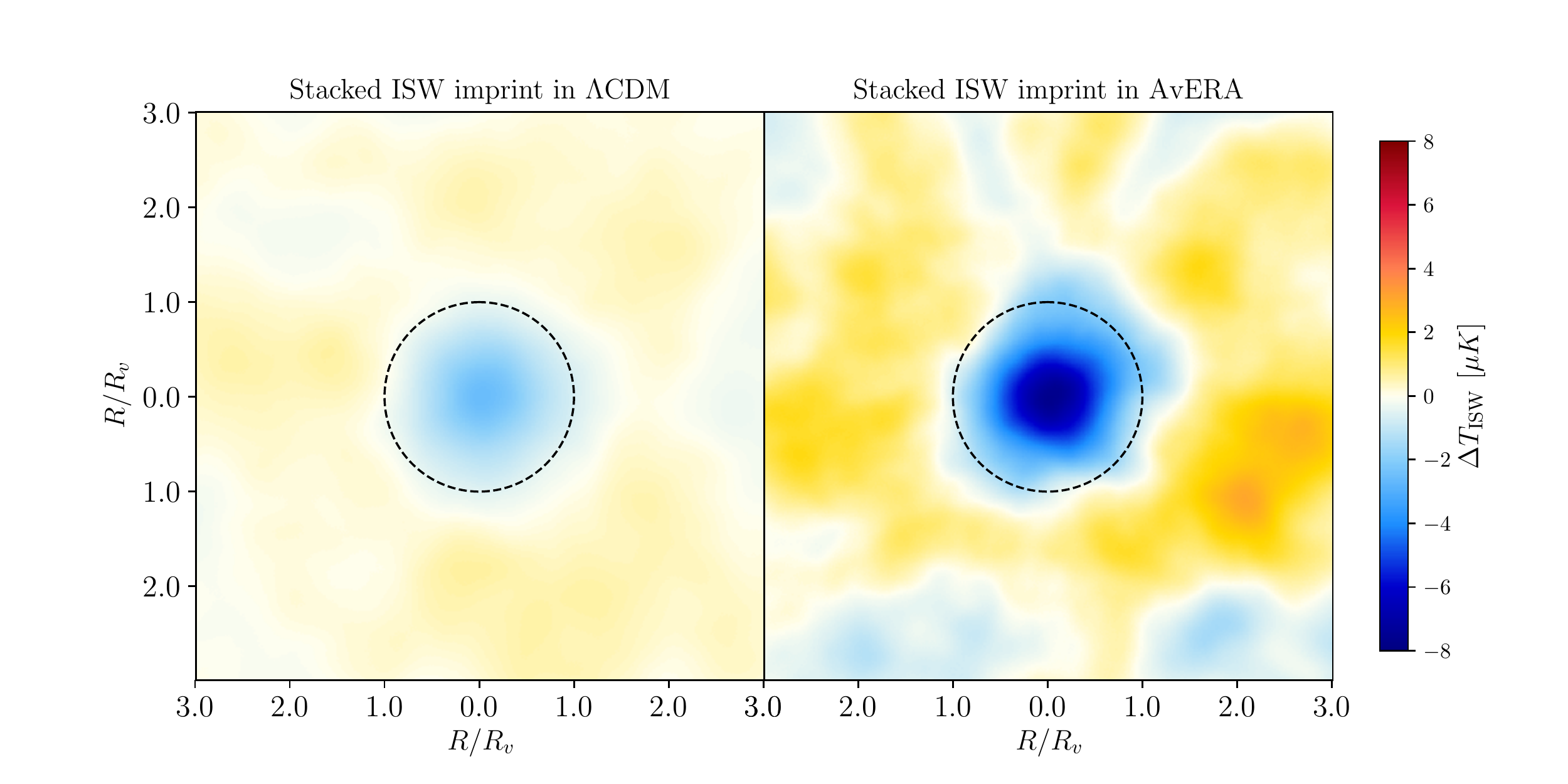}
\caption{\label{fig:figure_4}A comparison of stacked images of MXXL supervoids using $\Lambda$CDM (left) and AvERA (right) ISW maps. The data is presented with identical colour scales. The dashed circles mark the void radius in $R/R_{v}$ re-scaled radius units.}
\end{center}
\end{figure*}

We note that the ISW analysis without re-scaling, and using smaller voids, resembles more closely the traditional two-point correlation measurements than our main analysis. In fact, the signal-to-noise ratio expected from the stacking measurement of \cite{NadathurCrittenden2016} ($S/N\approx1.88$) is similar to what is expected from two-point functions applied to the same data sets ($S/N\approx1.79$). The main difference is that if the ISW imprint profile is binned in angular size $\theta$ without re-scaling (as in two-point function measurements), then edges of cosmic voids are not stacked on top of each other even though their centres overlap. Consequently, the true extent of large-scale structures is not taken into account and some phase information is lost, unlike when using $R/R_{v}$ relative angular size.

\section{Results}

\subsection{Stacked images of supervoids}

We now describe the characteristics of the stacked AvERA and $\Lambda$CDM ISW signals in MXXL. Then, we compare our findings to real-world data from the DES and BOSS surveys. 

In Figure \ref{fig:figure_4}, we show a simple comparison of the stacked AvERA and $\Lambda$CDM imprints using temperature maps without $2\leq \ell \leq10$ modes that we removed from the ISW maps, as explained in Section 2.1, in order to avoid possible biases. We find that, in the shape of their imprints, the images are similar with hot rings surrounding cold spots. The \emph{amplitude} of the signal, however, is higher in the case of the AvERA image, reflecting also the findings by \cite{Beck2018} on the enhanced auto-correlation signal. The data shows a visually compelling $\Delta T_{0} \approx -6.8\pm0.6~\mu K$ cold spot in the centre of the stacked AvERA image while the coldest regions in the $\Lambda$CDM version are of $\Delta T_{0} \approx -2.3\pm0.2 ~\mu K$, in accordance with previous results from the $\Lambda$CDM model \cite[see e.g.][]{Nadathur2012,Flender2013,Hernandez2013}.

\subsection{The imprint profile of supervoids}

From the stacked images, we also measure radial ISW profiles in re-scaled void radius units using bins of $\Delta (R/R_{v})=0.1$ up to $R/R_{v}=3$. We estimate the corresponding uncertainties from 500 random stacking measurements. Given the different ISW auto power spectra \cite{Beck2018} calculated for $\Lambda$CDM and AvERA models, we generated 500 realisations of ISW maps for both models using the \texttt{synfast} routine of \texttt{HEALPix}. We then used the MXXL catalogue of supervoids for stacking measurements on these uncorrelated maps to estimate the 'theoretical' errors of the profile reconstruction itself, i.e. not the measurement error using observed CMB data that would clearly dominate the uncertainties.

\begin{figure}
\begin{center}
\includegraphics[width=90mm]{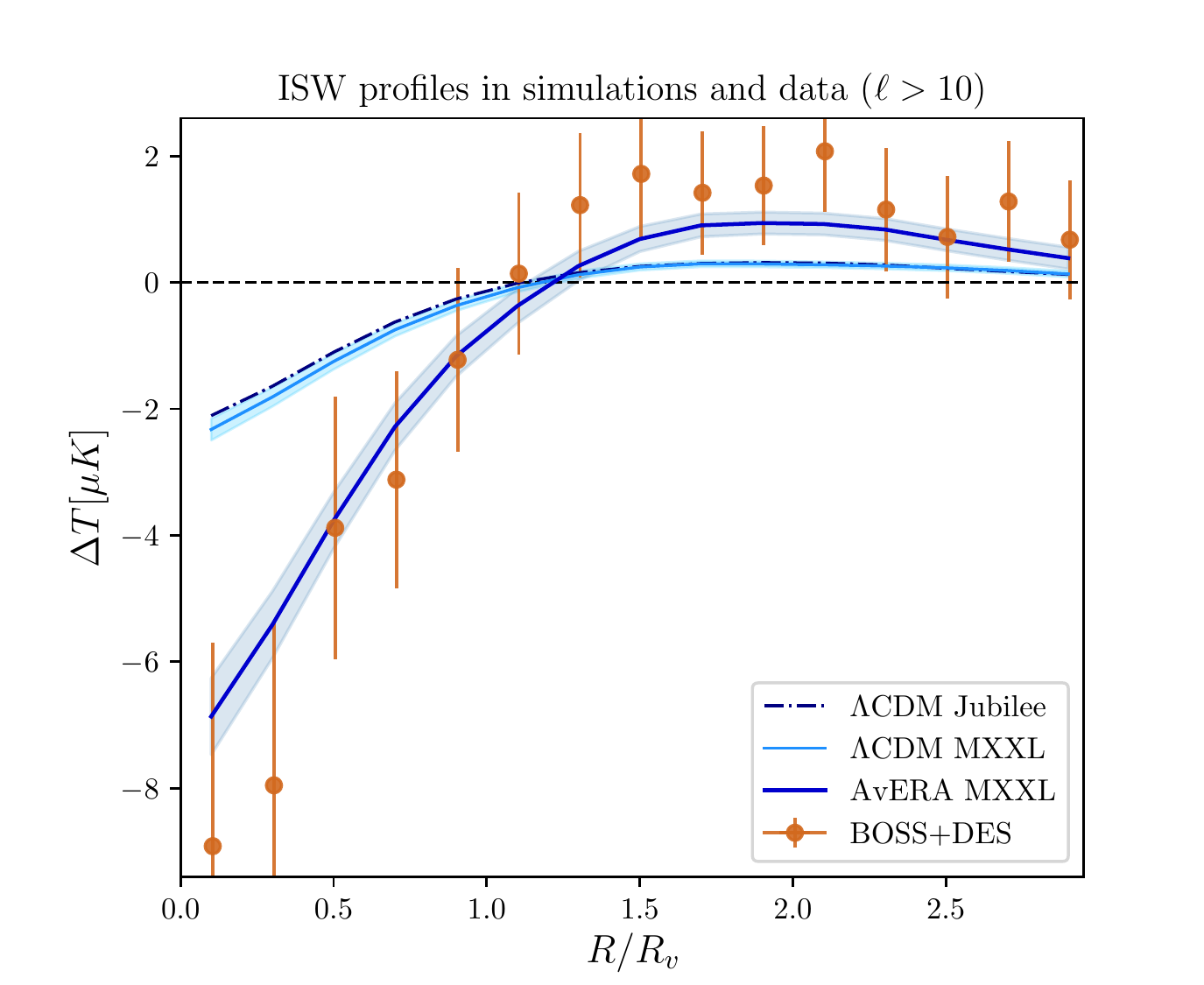}
\caption{\label{fig:figure_5}Radial profiles of ISW imprints in $\Lambda$CDM and in AvERA cosmologies are compared to observed ISW signals from a combined stacking analysis of DES and BOSS supervoids by \citet{Kovacs2019}. The shaded bands around simulated results show the standard deviation of profiles measured from random ISW-only stacking measurements. The BOSS+DES data points are shown with their actual observational error bars including CMB noise that dominates the uncertainties.}
\end{center}
\end{figure}

In our MXXL analysis, we find that the angular size of the cold spot in the centre does not distinguish the two cosmological models. The amplitude of the ISW effect, based on a simple comparison to the signal in the $\Lambda$CDM model  with $A_{\rm ISW}=\Delta T^{\rm obs} / \Delta T^{\rm \Lambda CDM}$, is a better measure of possible anomalies. \cite{Kovacs2019} reported a $3.3\sigma$ detection of an ISW signal from the combination of 87 DES and 96 BOSS supervoids with $A_{\rm ISW}\approx5.2\pm1.6$ amplitude, i.e. a $2.6\sigma$ tension in comparison with $\Lambda$CDM ($A_{\rm ISW}=1$). We found that the stacked signal from real-world DES and BOSS supervoids is consistent with the AvERA imprint, while it is in clear contrast with $\Lambda$CDM predictions as shown in Figure \ref{fig:figure_5}.

We consider the approximate agreement of simulations and data as a success of the AvERA prediction of the signal but do not attempt to fit the model to the data. Testing different AvERA versions, we found that the stacked ISW signal decreases on average by $\approx5\%$ using $N_{c}$=625,000 cells for coarse graining compared to our fiducial case with $N_{c}$=1,080,000. The observed data in fact favours a slightly higher amplitude than the AvERA result.

\begin{figure*}
\begin{center}
\includegraphics[width=175mm]{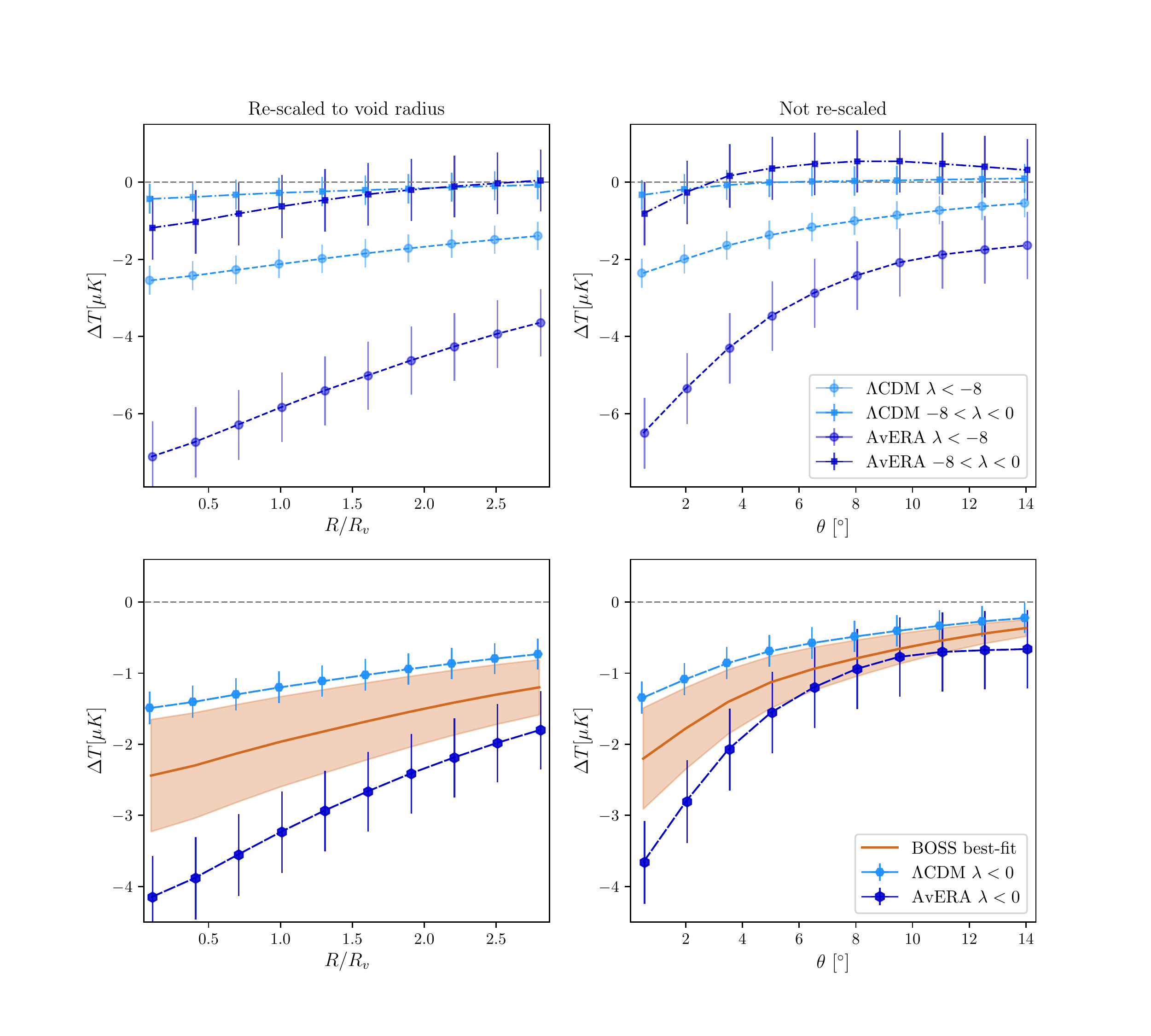}
\caption{\label{fig:figure_6}Imprints of \texttt{REVOLVER} voids in $\Lambda$CDM and AvERA models with re-scaling (left) and without re-scaling (right). In the top panels, void catalogs are split into two sub-groups with $-8<\lambda_v <0$ and $\lambda_v <-8$ in order to test void imprint properties in greater detail. The bottom panels illustrate a comparison of best-fit BOSS results with amplitude $A_{\rm ISW}\approx1.64\pm0.53$ (uncertainties marked by shaded brown areas) against $\Lambda$CDM and AvERA predictions for a combined sample of voids $\lambda_v <0$ following \citet{NadathurCrittenden2016}. The $\Lambda$CDM MXXL signal was scaled with a factor of 1.64 in order to make a comparison with AvERA predictions. Error bars are based on our simulated ISW map analyses, while shaded areas mark observational uncertainties on the ISW amplitude with $\sigma_{A_{\rm ISW}}\approx0.53$.}
\end{center}
\end{figure*}

\subsection{On the robustness of the ISW anomaly}

Testing the robustness of simulation methods and differences in input cosmology, we compare our MXXL result to those of \cite{Kovacs2019}. They estimated the ISW signal of simulated DES-like supervoids using a mock LRG tracer catalogue from the Jubilee N-body simulation with $6000^3$ particles that trace a volume of ($6h^{-1}$ Gpc)$^{3}$ based on a WMAP-5 $\Lambda$CDM cosmology \citep{Watson2014}. In Figure \ref{fig:figure_5}, we show that our MXXL estimate of the $\Lambda$CDM signal is in close agreement with the Jubilee results in the full extent of the profile. This agreement highlights that small differences in $\Lambda$CDM cosmological parameters or in simulation methodologies are not expected to explain the observed excess ISW signals.

On the observational side, we also try to answer why some approaches detect anomalous imprints, while others, using seemingly identical methods, do not observe excess signals from the very same data set. In particular, evidence exists for anomalous ISW-like signals associated with supervoids from repeated observations with moderate to high significance \citep[see e.g.][]{Granett2008,Cai2017,Kovacs2016,Kovacs2018,Kovacs2019}. Yet, critics consider these findings chance fluctuations even though they have been seen in independent data sets and non-overlapping parts of the sky. These objections are fair because the excess signals appear to contradict presumably more accurate measurements that are based on two-point cross-correlation or other void catalogues containing smaller voids \citep[see e.g.][]{PlanckISW2015, NadathurCrittenden2016, Stolzner2018}. Such inconsistencies are hard to interpret in the $\Lambda$CDM model or in typical alternative models \citep[e.g.][]{CaiEtAl2014}.

\subsection{\texttt{REVOLVER} voids and the observed ISW excess}

In the context of \texttt{REVOLVER} voids, non-detections of excess signals should also be understood in order to claim a viable solution to the ISW anomalies. In particular, a potentially important detail in the stacking procedure is whether re-scaling is applied to the stacked images, knowing the angular void size, or patches are stacked in absolute angular size. We thus measure the imprint of \texttt{REVOLVER} voids \emph{with and without} re-scaling in order to test any effect of such methodological details in the main results. Following \cite{NadathurCrittenden2016}, we stack on void positions using ISW maps with all available modes included ($\ell\geq2$).

We first split the pruned catalogue of 15,016 \texttt{REVOLVER} voids with $\lambda_v <0$ into two subgroups of similar size to study the sample in greater details (the lower the value of $\lambda_v$ the stronger the ISW imprint.) The most extreme 7,820 voids with $\lambda_v <-8$ form one subgroup, while 7,196 voids of $-8<\lambda_v <0$ constitute the second category with a smaller expected ISW amplitude. In Figure \ref{fig:figure_6}, we show that, as expected, $\lambda_v <-8$ voids imprint a significantly stronger signal than voids of $-8<\lambda_v <0$ in both models. The $\Lambda$CDM model shows a minimum central imprint $\Delta T_{0} \approx -2.5~\mu K$ while the AvERA model predictions are colder with $\Delta T_{0} \approx -6~\mu K$, regardless of re-scaling.

As a consistency check, we note that these $\Delta T_{0}$ values are in close agreement with the coldest central depressions found in our analysis of MXXL supervoids (see Figure \ref{fig:figure_5}). This finding indicates that the supervoid and \texttt{REVOLVER} approaches trace the same underlying large-scale patterns both in the tracer catalogue and in the ISW maps.\footnote{We again observed $\approx5\%$ smaller ISW signal if an alternative AvERA model version with $N_{c}$=625,000 cells is used.}

An important difference, however, is that the imprint of \texttt{REVOLVER} voids extends far beyond the actual re-scaled void size unlike in the case of supervoids that are built up from several of these smaller voids. This feature is a direct consequence of the lack of void merging and the relatively small resulting void size using this definition \citep[see e.g.][and references therein]{Kovacs2018}. 

We then combine the results and take the average of the two $\lambda_v$ subgroups in order to model the joint-fit measurement by \cite{NadathurCrittenden2016} who used all voids with $\lambda_v <0$. They reported a best-fit amplitude of $A_{\rm ISW}\approx1.64\pm0.53$ from their BOSS analysis, i.e. approximately $1.2\sigma$ higher than expected in $\Lambda$CDM ($A_{\rm ISW}=1$). We note that our tracers and voids may differ in details from theirs. Also, we do not apply their more advanced matched filtering analysis using several $\lambda_v$ bins of voids (and also superclusters) in order to fit the amplitude of the $\Lambda$CDM signal. We argue, however, that by simply scaling our MXXL $\Lambda$CDM imprint profile by a factor of their best-fit amplitude $1.64\pm0.53$ the results can be meaningfully compared to AvERA model predictions.

We stress that re-scaling was \emph{not} applied in their analysis. The consequence, as demonstrated in the right panels on Figure \ref{fig:figure_6}, is rather unimportant in the case of the $\Lambda$CDM model they assume. For instance, the central imprint of voids remains similar to re-scaled results at the $\Delta T_{0} \approx -1.3~\mu K$ level for the combined $\lambda_v<0$ sample. Measured profile values farther from the void centre, however, move closer to zero for $\lambda_v <-8$ voids or even become positive for the $-8<\lambda_v <0$ subgroup. Such behaviour is not unexpected because without re-scaling to the angular void size hot and cold signals from different parts of void profiles are averaged.

\begin{figure*}
\begin{center}
\includegraphics[width=180mm]{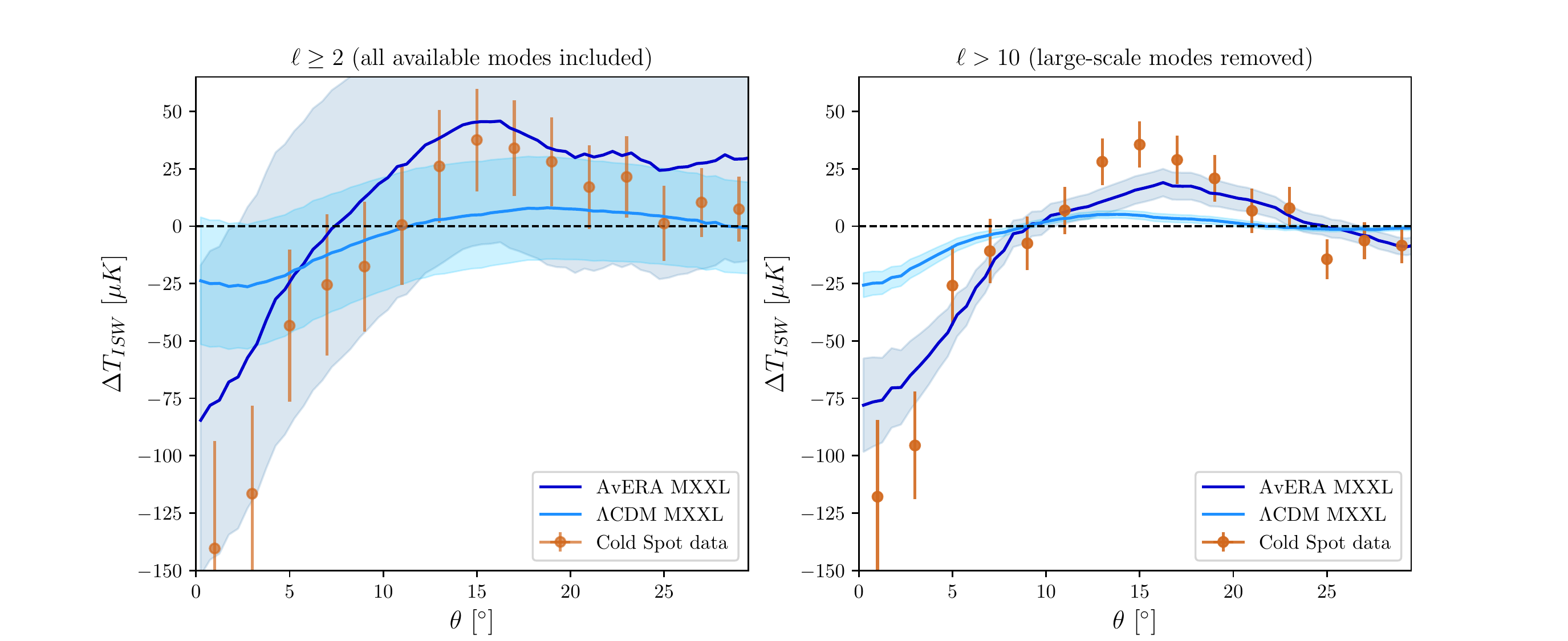}
\caption{\label{fig:figure_7}The coldest spots in AvERA and $\Lambda$CDM models are compared to observed profiles of the Cold Spot using all modes (left) and with large-scale modes removed (right). The shaded bands around simulated results show the standard deviation of profiles measured from ISW-only randoms, modelling theoretical errors on the estimated signals. The data points are shown with their realistic observational error bars including the dominating CMB noise.}
\end{center}
\end{figure*}

The imprints of \texttt{REVOLVER} voids in the AvERA model are more revealing. They appear to be more sensitive to whether re-scaling is applied because in AvERA not only the voids but also the surrounding over-densities leave stronger ISW imprints. Like in $\Lambda$CDM, the extreme $\lambda_v <-8$ voids show a consistent central minimum in AvERA regardless of the stacking strategy while data points in the outer parts of the imprint profile move closer to zero. Imprints of less extreme $-8<\lambda_v <0$ voids remain negative in the centre, but the mixing effect of contributions from different void parts are more pronounced near void boundaries. In AvERA, the lack of re-scaling thus results in a $\Delta T >0$ pattern in a large part of the outer profile where the $\Lambda$CDM version shows zero or slightly negative signal.

Most importantly, it is revealed that in a large part of the AvERA imprint profile this $\Delta T >0$ signal from less extreme $-8<\lambda_v <0$ voids partially \emph{cancels} the contributions from cold spots imprinted by more extreme $\lambda_v <-8$ voids when signals are averaged. Therefore, characteristic differences between $\Lambda$CDM and AvERA models are greatly reduced when the mean signal of all voids ($\lambda_v <0$) is considered without re-scaling.

These intriguing details should be interpreted in the light of observational constraints. Importantly, we found that while AvERA predictions exceed the range constrained by BOSS data at the $\sim2\sigma$ level in the void centre, the two imprints are generally consistent within errors in most of the profile, i.e. the otherwise detectable excess imprints from AvERA have become \emph{undetectable} without re-scaling.

Intriguingly, \cite{NadathurCrittenden2016} in fact also found colder-than-expected $\Delta T_{0}$ values in some of the most extreme $\lambda_v$ bins in their data. Yet, the more numerous and less noisy data points corresponding to more ordinary $\lambda_v$ bins possibly dominated their linear joint-fit procedure, leading to a general $A_{\rm ISW}\approx1.64\pm0.53$ best-fit amplitude. We note that \cite{Cai2017} and \cite{Kovacs2018} did find evidence for excess ISW signals considering different variants of \texttt{ZOBOV}-based void catalogues extracted from the same BOSS galaxy catalogue, and by applying re-scaling to void size in the stacking process.

We therefore argue that our findings can explain how an alternative cosmology's characteristics may be concealed depending on arbitrary choices in the data analysis, leading to seemingly inconsistent observational constraints on the ISW amplitude from competing methodologies.

\section{On the Cold Spot anomaly}

We now seek to stress-test a further prediction of the AvERA model at the extremes. In order to understand how the problem of \emph{the} Cold Spot in the CMB sky \citep[see e.g.][]{CruzEtal2004} may be related to the ISW puzzle, we compare AvERA predictions for the \emph{coldest} spot in an ISW map to its $\Lambda$CDM and observed equivalents as in previous examples.

\subsection{The supervoid in alignment}

There is ample evidence for the low-redshift Eridanus supervoid aligned the CMB Cold Spot from galaxy density maps \citep{FinelliEtal2014, SzapudiEtAl2014} and also from large-scale reconstructions of cosmic flows \citep{Courtois2017}. The supervoid of radius $R_{\rm v}\gsim 200 \mpc$ and central density $\delta_0 \approx-0.25$ appears to be elongated in the line-of-sight \citep{KovacsJGB2015} with several sub-voids \citep{Mackenzie2017}. 

There is also a consensus about the corresponding ISW imprint of model supervoids with parameters consistent with the above observational characteristics \citep[see e.g.][]{Nadathur2014,MarcosCaballero2015,Naidoo2016,Naidoo2017}. The expected central ISW imprint is of order $\Delta T_0 \approx -20~\mu K$, in accordance with the coldest spot in the Jubilee ISW simulation using the same definition \citep{Kovacs2018}. Then, among others, \cite{Nadathur2014} and \cite{Mackenzie2017} concluded that the Eridanus supervoid and the Cold Spot \emph{cannot} be in causal relation because the modelled ISW imprint of the void is not sufficient to explain the Cold Spot profile with $\Delta T_0 \approx -150~\mu K$. 

\begin{figure*}
\begin{center}
\includegraphics[width=185mm]{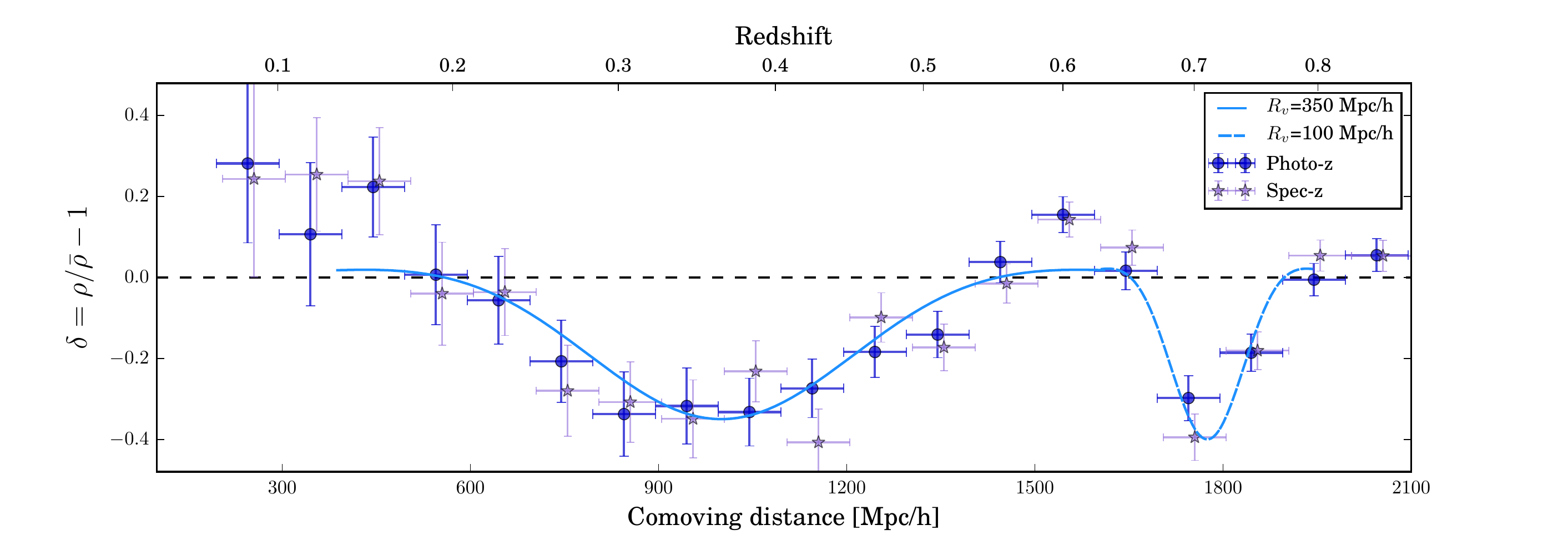}
\caption{\label{fig:figure_8} Measurements of matter density along the line-of-sight aligned with the coldest spot in the AvERA ISW map. In several tomographic slices of $100~\mpc$ width, we use a constant aperture radius of $\theta=2.5^{\circ}$ (i.e. the inner half of the $5^{\circ}$ cold spot, where the under-density is expected to be more pronounced). As expected for large supervoids from previous measurements using DES data and simulations, the density mapping remains consistent when we add DES-like photo-$z$ noise, or use accurately known tracer positions (spec-$z$) available in the MXXL data. We model the supervoids using a $\delta(r) =  \delta_0\, \Big( 1 - \frac{2}{3}\,\frac{r^2}{r_0^2}\Big)\,\exp\Big[\!-\frac{r^2}{r_0^2}\Big]\,$ profile following Finelli et al. (2014) and report that the data is consistent with two large supervoids centered at redshifts $z\approx 0.35$ and $z\approx 0.7$.}
\end{center}
\end{figure*}

Part of the sensible reasoning was that in the standard theory of peaks for Gaussian random fields \citep{BBKS1986} the probability of the formation of a supervoid capable of imprinting a Cold-Spot-like profile is practically zero \citep{Nadathur2014}. The Cold Spot itself is a $\sim3\sigma$ fluctuation in Gaussian CMB map statistics of cold spots and thus hypothesising such an unlikely supervoid makes no sense in solving the problem itself.

As in the case of the higher-than-expected stacked signals of supervoids observed elsewhere in the sky, however, critics of the Cold Spot's supervoid hypothesis presented a false dichotomy. The above studies all assumed an underlying concordance $\Lambda$CDM model to estimate these ISW signals, and then concluded that no causal relation is possible because of the disagreeing models and observation \citep[see e.g.][]{NadathurCrittenden2016}. One may argue, however, that the original detection of excess ISW signals by \cite{Granett2008} or the enhanced $A_{\rm ISW}\approx5.2\pm1.6$ amplitude of the more numerous DES and BOSS supervoids reported by \cite{Kovacs2019} both raise the possibility that there may be a causal relation in a hitherto unknown alternative cosmological model. 

\subsection{The coldest spot in MXXL}

Looking for the coldest spot in our simulated MXXL ISW maps, we use a Spherical Mexican Hat Wavelet (SMHW) filter that was originally used to detect the Cold Spot in WMAP data \citep[see e.g.][]{CruzEtal2004}. The SMHW filter is equivalent to a second derivative of a Gaussian, resembling the key characteristics of a central cold spot and a compensating hot ring, i.e. the shape of the ISW imprint of supervoids. 

In our analysis, we follow\footnote{A python code for calculations of the filtered signal was kindly provided by Naidoo et al.} \cite{Naidoo2017} and filter the MXXL ISW maps using an SMHW filter of radial scale $R=5^{\circ}$ to match the original detection criteria. Patches of the map with the characteristic cold-spot-plus-hot-ring profile are thus identified, and the most extreme patch with the most substantial response to the SMHW filter is selected as the coldest spot in the map. For further details about the filtering methodology see Appendix \ref{smhw_details}. 

We first identified the locations of the coldest spots in our filtered MXXL maps and then measured the ISW temperature profiles around those centres. Then measurement errors were estimated from the same set of 500 random realisations of the $\Lambda$CDM and AvERA ISW maps that we used in the stacking measurements. We present our findings in Figure \ref{fig:figure_7}.

In our main analysis, we first of all found that the removal of large-scale modes again plays a role. The coldest spot in the AvERA map with all modes included ($\ell\geq2$, Figure \ref{fig:figure_7}) shows no convergence to zero signal at large radii, indicating a bias due to an actual positive large-scale fluctuation it sits on. Nevertheless, the AvERA prediction again exceeds that of the $\Lambda$CDM model. While fluctuations are important, the observed Cold Spot profile is found to be consistent with the AvERA prediction given the errors and modulo the bias that we do not remove from the data.

More interestingly, the AvERA model receives even more support from the Cold Spot data in the analysis without large-scale modes in the maps ($\ell>10$, Figure \ref{fig:figure_7}). With reduced uncertainties, we observe that the AvERA prediction is in great agreement with the Cold Spot observations not just in terms of the size of the imprint, but also in the amplitude that the $\Lambda$CDM version cannot fit well. 

We highlight that, analogously to the case of the stacked signals of supervoids, the observed data appears to favour an even higher amplitude than that of the AvERA model. As in the case of stacked signals, we again observe a $\approx5\%$ smaller AvERA ISW signal using $N_{c}$=625,000 instead of the fiducial model thus the results are fairly robust. Key features of the profile such as the central cold spot, zero crossings, and the surrounding hot ring are all explained remarkably well by the AvERA framework.

In order to test the source of this prominent cold spot in the AvERA ISW map\footnote{Visible near the centre of the mollweide plot of the map in the bottom-right panel of Figure \ref{fig:figure_2}.}, we measured the mean matter density along the line-of-sight in its position. As illustrated in Figure \ref{fig:figure_8}, the MXXL data is consistent with intersecting two extended supervoids with radii $350~\mpc$ and $100~\mpc$ (both with central underdensity $\delta_{0}\approx-0.3$), centered at redshifts $z\approx0.35$ and $z\approx0.7$, respectively. A spherical supervoid of $R_{\rm v}\approx350~\mpc$ radius combined with $\delta_{0}\approx-0.3$ would be a rare (but not impossible) structure in a $\Lambda$CDM universe \citep[see e.g.][for details]{Nadathur2014}. However, the existence of the super-structure we detected is by definition consistent with the $\Lambda$CDM model because we identified it in the MXXL simulation. Therefore, a more plausible interpretation is an elongated structure built up from aligned $R_{\rm v}\approx100~\mpc$ supervoids that are fairly typical in our supervoid catalog (see Figure \ref{fig:figure_3}). Leaving any more detailed analyses for future work, we conclude that the rare supervoid structures we identified as sources of the coldest spot in the AvERA ISW map are comparable, both in size and in under-density, to the Eridanus supervoid that was found in alignment the observed Cold Spot in the CMB \citep{SzapudiEtAl2014,KovacsJGB2015,Mackenzie2017}.

\begin{figure}
\begin{center}
\includegraphics[width=90mm]{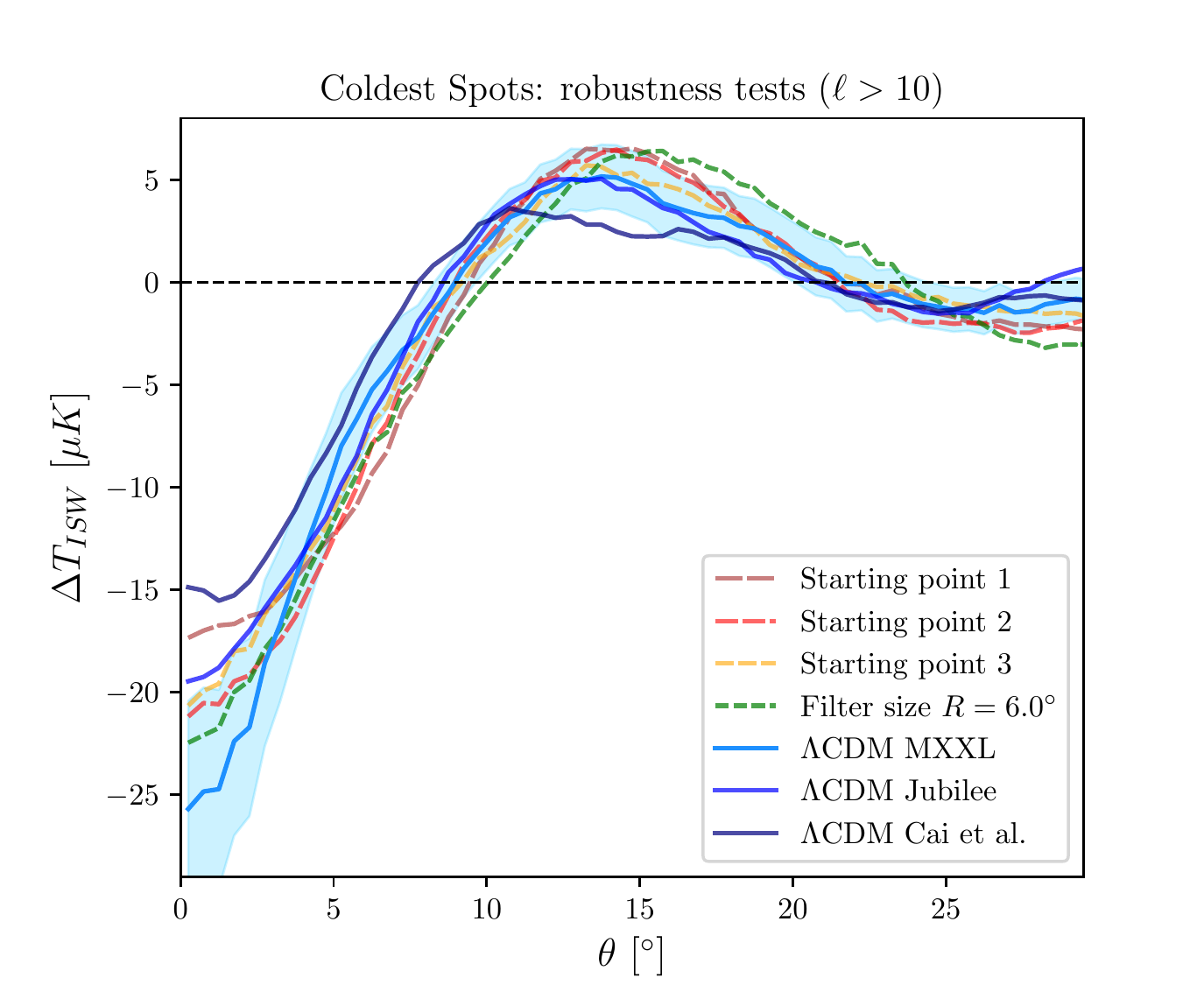}
\caption{\label{fig:figure_9} Comparison of the coldest SMHW-filtered spots in different $\Lambda$CDM simulations (MXXL, Jubilee, \citet{Cai2010}) and in different MXXL analysis setups (different starting points and altered filter scale).}
\end{center}
\end{figure}

Then, as a consistency check, we repeated the SMHW analysis using ISW maps from other $\Lambda$CDM N-body simulations \citep{Watson2014,Cai2010}. We found that, given the errors, the profiles of coldest spots are consistent in all data sets of slightly different $\Lambda$CDM parameters. We then also tested if the starting point of the ray-tracing method in the simulation box, or a slightly different filter scale may change the results. We again found no important differences and concluded that our MXXL analysis is robust. These findings are summarised in Figure \ref{fig:figure_9}.

We note that the shape of the Cold Spot has also been found consistent with profiles of the coldest spots in Gaussian CMB simulations. In fact, not its coldness but the combination with a surrounding hot ring makes it anomalous at the $\sim3\sigma$ level \citep{Nadathur2014}. We argue, however, that even though random CMB maps may produce similar cold spot features, the explaining power of the AvERA model should again be considered a success for models of emerging curvature in general, especially in the light of similar results for the stacked sample of supervoids.

\section{Discussion}

\subsection{$\Lambda$CDM anomalies and the AvERA model}

We explored how the explaining power of a possible solution of the $H_{\rm 0}$ tension may generalise to interpret other observational anomalies of the $\Lambda$CDM model. We argued, following \cite{Beck2018}, that a seemingly unrelated and less studied \emph{late-time} anomaly of the $\Lambda$CDM model --the higher-than-expected ISW imprint of cosmic super-structures-- is plausibly related to the $H_{\rm 0}$ problem in the so-called AvERA model of emerging curvature \citep{Racz2017}. 

The key character of the AvERA approach is the replacement of the cosmological constant ($\Lambda$) with a backreaction effect of large-scale inhomogeneities that may be interpreted as an emerging curvature term \citep[see e.g.][for critical views and relations to cosmological backreaction]{Buchert2000,Wiltshire2009,Rasanen2011,Buchert2012,Kaiser2017,Roukema2018,Buchert2018}. The result, in a modified Einstein--de Sitter setting with $\Omega_{\rm M}=1$ initially, is an expansion history consistent with that of the $\Lambda$CDM model, i.e. the observation of apparent late-time cosmic acceleration is not questioned.

As a consequence, the AvERA model predicts larger negative values of the derivatives of the growth of structure at low redshifts, directly leading to an enhanced amplitude for the ISW effect that can discriminate it from $\Lambda$CDM. In pursuit of this characteristic ISW signal, we built on the related ray-tracing ISW analysis and auto-correlation measurements by \cite{Beck2018} in the MXXL simulation, but formulated different questions:
\begin{enumerate}
\item How the AvERA imprints compare to $\Lambda$CDM estimates given different void definitions including supervoids?
\item Do methodological details like mapmaking, filtering, or stacking strategy play some role in the results?
\item How the AvERA and $\Lambda$CDM imprints compare to observed excess signals from supervoids?
\item How the CMB Cold Spot anomaly may be related to the problem?
\item Can our related tests confirm the applicability of the AvERA model as a resolution of the $H_{\rm 0}$ tension?
\end{enumerate}

\subsection{ISW analyses using the MXXL simulation}

We performed a comprehensive analysis of simulated supervoids  and their corresponding ISW imprints in the MXXL simulation. We relied on a public light-cone halo catalogue for finding the largest under-densities at $0.2 <z<0.9$. We then used a ray-traced reconstruction of the ISW signal to stack cut-out patches in the positions of supervoids given $\Lambda$CDM and AvERA cosmologies. 

We found significantly stronger imprints in the case of the AvERA model. The estimated signals are perfectly consistent with the excess imprint observed from a sample of 183 supervoids, mapped by the DES and BOSS galaxy surveys, that have previously been difficult to explain in the $\Lambda$CDM model \citep[see e.g.][]{Nadathur2012}. 

However, the explaining power of the AvERA model in the case of supervoids is not yet sufficient to jump to conclusions because the overall state-of-the-art $\Lambda$CDM view of the ISW problem, taken at face value, is the following:
\begin{itemize}
\item[$-$] \cite{NadathurCrittenden2016} reported that there is no significant ISW tension given the $A_{\rm ISW}\approx1.64\pm0.53$ best-fit ISW amplitude they constrained by using a populous catalogue of BOSS voids (and also superclusters).
\item[$-$] among other studies, they reported that their results are in contrast with the excess signals by \cite{Granett2008} and thus, given their presumably more precise measurements, the latter must be a rare fluctuation.
\item[$-$] finally, they also concluded that the lack of a very large enhancement of the $A_{\rm ISW}$ amplitude invalidates the supervoid explanation of the CMB Cold Spot, again suggesting a chance alignment with a supervoid.
\end{itemize}

While the recently identified excess ISW signals \emph{elsewhere} in the sky from the Dark Energy Survey Year-3 data with $A_{\rm ISW}\approx5.2\pm1.6$ amplitude (in combination with BOSS results using supervoids instead of smaller voids) may suggest that the problem is real, the above high-precision findings cannot be bypassed. We therefore tested the possible role of the definition of voids since while \cite{NadathurCrittenden2016} used rather small voids (and superclusters), excess signals have appeared using supervoids that trace larger scales in the cosmic web by definition.

Modelling their BOSS CMASS measurements at $0.4<z<0.7$ in our MXXL framework, we reported that seemingly unimportant methodological differences may lead to important differences in the outcomes. In particular, a stacking measurement \emph{without} re-scaling, advocated among others by \cite{NadathurCrittenden2016}, covers up the distinct signatures of the AvERA model almost entirely. Effectively, their slightly enhanced $A_{\rm ISW}\approx1.64\pm0.53$ best-fit value is in fact consistent with the AvERA prediction throughout most of the profile because of a mixing effect of cold imprints with the enhanced hot ring parts of the signal at void edges. These effects are the most important for the less extreme voids that are also the most numerous and thus may dominate the inference on the ISW amplitude.

Finally, we also investigated the case of \emph{the} CMB Cold Spot by finding the equivalent coldest spots in the AvERA and $\Lambda$CDM versions of the MXXL simulation. 
We again reconstructed an ISW profile in the AvERA model that is consistent with {\it Planck} observations of the Cold Spot, i.e. significantly larger than expected in the $\Lambda$CDM model. Simulation and observation agree remarkably well in key features of the profile such as the central temperature, locations of zero crossings, and the size and magnitude of the surrounding hot ring.

\section{Conclusions}

In summary, we used the AvERA model to make the following three statements about the ISW puzzle:

\begin{itemize}
\item the observed excess ISW signal of real-world supervoids can accurately be modelled in the AvERA cosmology.
\item perfectly valid but different definitions of voids paired with specific methodological details can obscure the otherwise detectable AvERA ISW signal.
\item the observed profile of \emph{the} CMB Cold Spot resembles closely the coldest spot in the AvERA ISW map.
\end{itemize}

All things considered, we report that supervoids at $\gsim100\mpc$ scales are particularly sensitive to differences between $\Lambda$CDM and AvERA cosmologies. Our results suggest that the recalcitrant evidence for an $A_{\rm ISW}\approx5.2\pm1.6$ excess signal from supervoids, an $A_{\rm ISW}\approx1.64\pm0.53$ best-fit ISW amplitude using smaller void types, and a $\Delta T_0 \sim -100~\mu K$ cold spot of ISW origin in the CMB sky can all be consistent within a given cosmological model. The difference in our interpretation, however, is that this model is not the $\Lambda$CDM model in which the ISW signal of supervoids is considered a rare fluctuation in an otherwise well-functioning model. Instead, we argue that the AvERA model can offer a solution not just for the $H_0$ tension but, without tuning any free parameter, also for the ISW excess signal of super-structures. Observed ISW imprints, therefore, may become from solid evidence for dark energy to evidence against it in favour of the emerging curvature family of models.

The prevailing view is that the Friedmann-Robertson-Walker (FRW) expansion rate is influenced by inhomogeneities, while there is debate on the magnitude of the effect. Some, based on simple Newtonian considerations \citep[e.g.][]{Kaiser2017}, or GR simulations \citep{Macpherson2019}, claim that such backreaction is negligible. Others, through non-linear approximations and averaging of Einstein's equations \citep[see e.g.][]{Buchert2012,Bolejko2018}, argue for emerging curvature. We consider this question unsettled. It remains to be resolved with new theoretical and observational research, and possible re-interpretation of existing observations in new and more sophisticated models of backreaction effects that AvERA may approximate \citep[see e.g.][for further discussion]{Wiegand2010,Heinesen2020}.

As an outlook, we mention that the AvERA model predicts a slightly faster gravitational growth at $z \approx 1.5 - 4.4$ before the emerging curvature freezes progress even faster than $\Lambda$CDM. The resulting transient growth function is a smoking gun of AvERA, most likely shared with other emerging curvature models, and it gets around sound horizon problems of the class of monotonic late solutions to the Hubble tension. In the near future, the opposite-sign ISW effect at $z \approx 1.5 - 4.4$ may also be tested with galaxy surveys like DESI or Euclid beyond simple follow-up studies on low-$z$ ISW anomalies AvERA resolves. Furthermore, the gravitational lensing signal of voids is yet another observational probe that may discriminate between AvERA and $\Lambda$CDM models \citep{Cai2017,Vielzeuf2019,Raghunathan2019}.

\section*{Acknowledgments}

The authors thank Julian Moore, Francisco-Shu Kitaura, and Juan Garc\'ia-Bellido, and the anonymous reviewer for insightful discussions about the results presented in the paper. AK has been supported by a Juan de la Cierva fellowship from MINECO with project number IJC2018-037730-I. IS and RB acknowledge support from the National Science Foundation (NSF) award 1616974, and from the National Research, Development and Innovation Office of Hungary via grant OTKA NN 129148. Funding for this project was also available in part through SEV-2015-0548 and AYA2017-89891-P. 

\section*{Data availability}
The data and software underlying this article are available from public websites specified in various footnotes, or will be shared on reasonable request to the corresponding author.

\appendix

\section{Ray-tracing details}
\label{app_raytracing}

We repeat the methodological details in ray-tracing map reconstruction techniques by \cite{Beck2018} below. We note that they used three different \emph{random} starting points for the ray-tracing in MXXL to produce ISW maps. In the present analysis, we instead use the centre of the simulation in the ray-tracing process in order to ensure density-ISW correlation following \cite{Smith2017} who positioned the observer into the centre.

The derivative of the $\Phi$ gravitational potential and the corresponding temperature shift $\Delta T_{\mathrm{ISW}}$ of the CMB are related using the integral
\begin{equation}
\frac{\Delta T_{\mathrm{ISW}}}{T_{\mathrm{CMB}}} = - \frac{2}{c^2} \int_{\tau_0}^{\tau_{\mathrm{CMB}}} d\tau \frac{d\Phi\left(\vec{x}(\tau),\tau\right)}{d\tau}
\label{eq:ISWintegral}
\end{equation}
over the path of a CMB photon, from present conformal time $\tau=\tau_0$ to the surface of last scattering where $\tau=\tau_{\mathrm{CMB}}$ \citep[e.g.][]{Sachs1967,Seljak1996,Cai2009,Cai2010}. $T_{\mathrm{CMB}}$ is the mean temperature of the CMB, $c$ is the speed of light, and $\vec{x}$ denotes comoving coordinates. 

The gravitational potential is related to density fluctuations via the Poisson equation
\begin{equation}
\vec{\nabla}^2\Phi\left(\vec{x},\tau\right)=4\pi G \overline{\rho}(\tau)a^2(\tau) \delta \left(\vec{x},\tau \right),
\end{equation} 
where $G$ is the gravitational constant, $\overline{\rho}$ is the average density of the universe, $a$ is the scale factor and $\delta=(\rho-\overline{\rho}) / \overline{\rho}$ is the density contrast. The Poisson equation in Fourier space reads as 
\begin{equation}
\Phi\left(\vec{k},\tau\right)=-\frac{3}{2} H_0^2 \Omega_m \frac{1}{a(\tau)} \frac{\delta\left(\vec{k},\tau\right)}{k^2},
\end{equation}
with a substitution $4 \pi G \overline{\rho} = \frac{3}{2} H_0^2 \Omega_m$, introduction the $H_0$ Hubble constant and the $z=0$ value of the matter density parameter, $\Omega_m$. The continuity equation in Fourier space, $\dot{\delta} + i\vec{k}\vec{p}=0$, where the overdot represents derivative with respect to cosmic time, is then used to express the derivative of the potential as
\begin{equation}
\frac{d\Phi\left(\vec{k}(\tau),\tau\right)}{d\tau} = \frac{3}{2} \frac{H_0^2}{k^2}\Omega_m \left[ H(\tau)\delta\left(\vec{k},\tau\right) + i \vec{k}\vec{p}\left(\vec{k},\tau \right) \right],
\label{eq:potderiv}
\end{equation}
where $\vec{p}\left(\vec{k},\tau\right)$ is the Fourier transform of $\vec{p}\left(\vec{x},\tau\right)=\left(1+\delta(\vec{x},\tau)\right)\vec{v}\left(\vec{x},\tau \right)$ --- the momentum density divided by the mean mass density ---, and $H(\tau)=\dot{a}/a$ is the Hubble parameter. 

Without momentum information, Eq.~\ref{eq:potderiv} for the derivative of the potential cannot be fully computed. As an approximation, linear growth can be assumed, $\delta\left(\vec{k},\tau\right)=D(\tau)\delta\left(\vec{k},\tau=\tau_0\right)$, using the linear growth factor $D(\tau)$. Replacing the continuity equation form of the derivative of $\delta$ with this approximation we get
\begin{equation}
\frac{d\Phi\left(\vec{k}(\tau),\tau\right)}{d\tau} = \frac{3}{2} \frac{H_0^2}{k^2}\Omega_m \left[ H(\tau)\delta\left(\vec{k},\tau\right) \left(1-\beta(\tau)\right) \right],
\label{eq:potderivlin}
\end{equation}
where we introduced $\beta(\tau) \equiv d \ln D / d \ln a $. We note that in a homogeneous Einstein--de Sitter $\mathrm{CDM}$ cosmological model, without a $\Lambda$ term, there is no ISW effect since $\beta(\tau)=1$. However, in any cosmological model with $\beta(\tau) \neq 1$ growth rate, there will be an ISW effect --- such as in AvERA, where there is no cosmological constant in the model, but the transient growth factor evolution is instead sourced by the treatment of inhomogeneities.

Therefore, in a given cosmological model, specified by the Hubble parameter and the growth factor, the ISW effect may be computed from a known density field by tracing the path of light through that field using Eqs.~\ref{eq:ISWintegral} and \ref{eq:potderivlin}. Then, the resulting ISW map may be transformed into a $C_l^{\mathrm{ISW}}$ spherical auto-correlation power spectrum using a harmonic analysis.

\cite{Beck2018} noted that, since MXXL is a $\Lambda \mathrm{CDM}$ simulation with an expansion history  slightly different from that of AvERA, the simulated matter density field cannot be directly paired to its AvERA equivalent. A reasonable solution, argued \cite{Beck2018} is that an early snapshot of the simulation can be considered when inhomogeneities are still small and different cosmologies match closely. This early state of the density field is then evolved, following the growth function and applying the linear growth approximation. Note that the linear growth approximation cannot be avoided since only density data is available for the MXXL snapshots. We therefore follow \cite{Beck2018} and use the first Millennium XXL snapshot within the redshift coverage of the AvERA run, at $z=8.55$, and ray-trace the corresponding simulation box to estimate the ISW effect. We repeat this procedure for both the $\Lambda \mathrm{CDM}$ and AvERA cosmologies to allow statistical comparisons and to facilitate out stacking measurements.

We then calculated Eq.~\ref{eq:ISWintegral} via Eq.~\ref{eq:potderivlin} by assuming linear growth. From the centre of the simulation box as a starting position, we projected light-rays ,,into the past'' in $12 \times 64^2$ directions, matching the \texttt{HEALPix} spherical coordinates with $\mathrm{NSIDE}=64$, up to $z=8.55$. We thus evaluated the derivative of the gravitational potential at locations spaced $0.75 \, \mathrm{Mpc}$ apart (in comoving coordinates) using a trilinear interpolation within grid points. Then, local contributions were numerically integrated along each light-ray to create an ISW temperature map for both $\Lambda \mathrm{CDM}$ and AvERA cosmologies.

We note that the evolution of the AvERA simulation showed an expansion history similar to that of the standard $\Lambda \mathrm{CDM}$ model. However, small characteristic differences have been found in $H(z)$ at the epoch of recombination and at redhsift $z=0$ compared to the $\Lambda \mathrm{CDM}$ model predictions. \cite{Racz2017} reported that these trends were sensitive to the particle mass which is a ,,free'' input parameter of AvERA (may be interpreted as the mass of gravitationally independent mini-universes). If initial conditions of the simulation were set up assuming a Planck best-fit $\Lambda \mathrm{CDM}$ cosmological model \citep{Planck2018}, a particle mass of $1.17 \times 10^{11} M_\odot$ provided a present-time Hubble parameter of $H_0=73.1 \Hunit$, matching the local measurement of \citep{Riess2016}, thus resolving the tension between local and CMB-based measurements of the Hubble parameter \citep[see][for more details]{Racz2017}. We also note, however, that the AvERA model was found less sensitive to the particle mass parameter in the ISW map reconstruction process as presented by \cite{Beck2018}.

\section{SMHW filtering}
\label{smhw_details}

In the Spherical Mexican Hat Wavelet (SMHW) filtering methods, we follow \cite{Naidoo2017}, and use a filter determined by its angular scale $R$ as:
\begin{equation}
	\Psi(\theta;R) = A_{wav}(R)\left(1+\left(\frac{y}{2}\right)^{2}\right)^{2}\left(2-\left(\frac{y}{R}\right)^{2}\right)\exp\left(-\frac{y^{2}}{2R^{2}}\right),
\end{equation}
where $y\equiv 2\tan(\theta/2)$ and $\theta$ is the angular separation between two points, $\hat{n}$ and $\hat{n}'$, on a sphere. The $A_{wav}(R)$ normalisation constant is defined as:
\begin{equation}
	A_{wav}(R) = \left[2\pi R^{2}\left(1+\frac{R^{2}}{2}+\frac{R^{4}}{4}\right)\right]^{-1/2}.
\end{equation}
The SMWH-filtered temperature, i.e. the value of a point at $\hat{n}$ as the SMHW transformation is applied to a sky patch with an angular radius of $\theta$, is given by:
\begin{equation}
\label{deltat}
	\Delta T_{wav}(\theta;\hat{n},R) = \int_{0}^{\theta}\Delta T(\vec{n}')\Psi(\theta';R)d\Omega',
\end{equation}
where $\hat{n}'$ are pixels located within an angular distance $<\theta$ from direction $\hat{n}$. The \texttt{query\_disc} \texttt{HEALPix} function is used to define such pixels. The SMHW of a single pixel, $\Delta \mathcal{T}_{\Psi}(\hat{n})$, is then calculated by integrating equation \ref{deltat} across the whole sky or up to an angular radius of $\theta \simeq 4R$ (since $\Psi\left(\theta\gtrsim4R;R\right)\simeq0$):
\begin{equation}
\label{smhw}
	\Delta \mathcal{T}_{\Psi}(\hat{n}) = \Delta T_{wav}(\pi; \hat{n},R) \simeq
	 \Delta T_{wav} (4R; \hat{n},R).
\end{equation}

See also \cite{CruzEtal2004} for original applications of this technique to CMB temperature maps in order to detect the Cold Spot.

\bibliographystyle{mnras}
\bibliography{refs}
\end{document}